\pdfoutput=1
\RequirePackage{ifpdf}
\ifpdf 
\documentclass[pdftex]{sigma}
\else
\documentclass{sigma}
\fi

\usepackage{bbm}
\numberwithin{equation}{section}
\DeclareMathOperator{\Pol}{Pol}
\DeclareMathOperator{\Ad}{Ad}
\DeclareMathOperator{\ad}{ad}
\DeclareMathOperator{\const}{const}
\DeclareMathOperator{\sgn}{sgn}
\DeclareMathOperator{\tr}{tr}
\newcommand{\dd}{\textrm{d}}
\newcommand{\tinyS}{\textrm{\tiny S}}
\newcommand{\tinyD}{\textrm{\tiny D}}
\newcommand{\tinyFLM}{\textrm{\tiny FLM}}
\newcommand{\overstar}{\star\!\!\!\!}
\newcommand*{\rsprod}[2]{\overrightarrow{\prod_{#1}^{#2}}\!\!\!\!\!\!\!\star\ }

\begin{document}

\newcommand{\arXivNumber}{1401.5819}

\allowdisplaybreaks

\renewcommand{\thefootnote}{$\star$}

\renewcommand{\PaperNumber}{067}

\FirstPageHeading

\ShortArticleName{Asymptotic Analysis of the Ponzano--Regge Model with NC Metric Boundary Data}

\ArticleName{Asymptotic Analysis of the Ponzano--Regge Model\\
with Non-Commutative Metric Boundary Data\footnote{This paper is a~contribution to the Special Issue on Deformations of
Space-Time and its Symmetries.
The full collection is available at \href{http://www.emis.de/journals/SIGMA/space-time.html}
{http://www.emis.de/journals/SIGMA/space-time.html}}}

\Author{Daniele ORITI~$^\dag$ and Matti RAASAKKA~$^\ddag$}

\AuthorNameForHeading{D.~Oriti and M.~Raasakka}

\Address{$^\dag$~Max Planck Institute for Gravitational Physics (Albert Einstein Institute),\\
\hphantom{$^\dag$}~Am M\"uhlenberg 1, 14476 Potsdam, Germany}
\EmailD{\href{mailto:daniele.oriti@aei.mpg.de}{daniele.oriti@aei.mpg.de}}

\Address{$^\ddag$~LIPN, Institut Galil\'ee, CNRS UMR 7030, Universit\'e Paris 13, Sorbonne Paris Cit\'e,\\
\hphantom{$^\ddag$}~99 av.\ Clement, 93430 Villetaneuse, France}
\EmailD{\href{mailto:matti.raasakka@lipn.univ-paris13.fr}{matti.raasakka@lipn.univ-paris13.fr}}

\ArticleDates{Received February 04, 2014, in f\/inal form June 14, 2014; Published online June 26, 2014}

\Abstract{We apply the non-commutative Fourier transform for Lie groups to formulate the non-commutative metric
representation of the Ponzano--Regge spin foam model for 3d quantum gravity.
The non-commutative representation allows to express the amplitudes of the model as a~f\/irst order phase space path
integral, whose properties we consider.
In par\-ticular, we study the asymptotic behavior of the path integral in the semi-classical limit.
First, we compare the stationary phase equations in the classical limit for three dif\/ferent non-commutative structures
corresponding to the symmetric, Duf\/lo and Freidel--Livine--Majid quantization maps.
We f\/ind that in order to unambiguously recover discrete geometric constraints for non-commutative metric boundary data
through the stationary phase method, the deformation structure of the phase space must be accounted for in the
variational calculus.
When this is understood, our results demonstrate that the non-commutative metric representation facilitates a~convenient
semi-classical analysis of the Ponzano--Regge model, which yields as the dominant contribution to the amplitude the
cosine of the Regge action in agreement with previous studies.
We also consider the asymptotics of the ${\rm SU}(2)$ $6j$-symbol using the non-commutative phase space path integral for the
Ponzano--Regge model, and explain the connection of our results to the previous asymptotic results in terms of coherent
states.}


\Classification{83C45; 81R60; 83C27; 83C80; 81S10; 53D55}

\renewcommand{\thefootnote}{\arabic{footnote}} 
\setcounter{footnote}{0}

\section{Introduction}

Spin foam models have in recent years arisen to prominence as a~possible candidate formulation for the quantum theory of
spacetime geometry
(see~\cite{P1} for a~thorough review). Their formalism derives mainly from topological quantum f\/ield
theories~\cite{B}, Loop Quantum Gravity~\cite{Rovellibook,Thiemannbook} and discrete gravity, e.g., Regge
calculus~\cite{Regge}.
On the other hand, spin foam models may also be seen as a~generalization of matrix models for 2d quantum gravity via
group f\/ield theory~\cite{F, O}.
For 3d quantum gravity, the relation between spin foam models and canonical quantum gravity has been fully cleared up.
In particular, it is known that the Turaev--Viro model~\cite{TuraevViro} is the covariant version of the canonical
quantization (\`a~la Witten~\cite{ReshetikhinTuraev, Witten}) of 3d Riemannian gravity with a~positive cosmological
constant, while the Ponzano--Regge model is the limit of the former for a~vanishing cosmological
constant~\cite{AGN, Mizoguchi91,KarimAlex}
(see also~\cite{Noui:2011im,Noui:2011aa, Pranzetti:2014} on incorporating the cosmological constant in 3d LQG
and~\cite{Sahlmann,Sahlmann:2011rv} for further work on relating 3d gravity to Chern--Simons theory and quantum group
structures). In this case, the spin foam 2-complexes have been rigorously shown to arise as histories of LQG spin
network states, as initially suggested in~\cite{Reisenberger96}, while the correspondence between LQG states and the
Ponzano--Regge boundary data had been already noted in~\cite{Rovelli:1993kc}.
However, in 4d the situation is less clear.
Several dif\/ferent spin foam models for 4d Riemannian quantum gravity have been proposed in the literature, such as the
Barrett--Crane model~\cite{BO2, BarrettCrane}, the Freidel--Krasnov model~\cite{Freidel07}, a~model based on the f\/lux
representation~\cite{BO3}, and one based on the spinor representation~\cite{EteraMaite}, while in the Lorentzian case
the Engle--Pereira--Rovelli--Livine model~\cite{Engle07b, Engle07a} represents essentially the state of the art
(see also~\cite{P2} for a~review of the new 4d models). These 4d models dif\/fer specif\/ically in their implementation
of the necessary simplicity constraints on the underlying topological BF theory, which should impose geometricity of the
2-complex corresponding to a~discrete spacetime mani\-fold and give rise to local degrees of freedom.
Thus, a~further study of the geometric content of the dif\/ferent spin foam models is certainly welcome.
In particular, one might hope to recover discrete Regge gravity in the classical limit of the model, since this would
imply an acceptable imposition of the geometric constraints at least in the classical regime.
Moreover, classical general relativity can be obtained from the Regge gravity by further taking the continuum limit,
which allows for some conf\/idence that continuum general relativity may be recovered also from the continuum limit of
the full quantum spin foam model.
The Regge action is indeed known to arise as the stationary phase solution in the 3d case in the large-spin limit for
handlebodies~\cite{DGH,KaminskiSteinhaus13}.
In 4d, Regge action was recovered asymptotically f\/irst for a~single 4-simplex~\cite{BDFHP} and later for an arbitrary
triangulation with a~f\/ixed spin labeling, when both boundary and bulk spin variables are scaled to
inf\/inity~\cite{CF,Han13a,Han13,HZ1,HZ2}.
Recently, in~\cite{HK,Hellmann13}, an asymptotic analysis of the full 4d partition function was given using microlocal
analysis, which revealed some worrying accidental curvature constraints on the geometry of several widely studied 4d
models.
This work considered only the strict asymptotic regime of the spin variables, without further scalings of the parameters
of the theory.
The work of~\cite{Han13b, Magliaro11} on the other hand dealt with the large-spin asymptotics of the EPRL model
considering also scaling in the Barbero--Immirzi parameter, with interesting results.
In particular, the analysis of~\cite{Han13b} used also the discrete curvature as an expansion parameter and identif\/ied
an intermediate regime of large spin values (dependent on the Barbero--Immirzi parameter) that seems to lead to the
right Regge behavior of the amplitudes in the small curvature approximation.

Classically, spin foam models, as discretizations of continuum theories, are based on a~phase space structure, which is
a~direct product of cotangent bundles over a~Lie group that is the structure group of the corresponding continuum
principal bundle (e.g., ${\rm SU}(2)$ for 3d Riemannian gravity)\footnote{In this paper, we are concerned exclusively with the
case of topological spin foam models with vanishing cosmological constant.
For non-topological models, such as 4d quantum gravity models, the physical conf\/iguration space is a~homogeneous
subspace (or, including the Barbero--Immirzi parameter, a~more general subspace) of a~Lie group, instead of a~Lie group.
Likewise, for a~non-vanishing cosmological constant, the conf\/iguration space is a~quantum group.
Therefore, in these cases the structure of the physical phase space is, strictly speaking, more involved than what is
implied above.}.
The group part of the product of cotangent bundles thus corresponds to discrete connection variables on a~triangulated
spatial hypersurface, while the cotangent spaces correspond to discrete metric variables (e.g., edge vectors in~3d, or
face bivectors in 4d, which correspond to discrete tetrad variables due to the simplicity constraints).
Accordingly, the geo\-metric data of the classical discretized model is transparently encoded in the cotangent space
variables.
However, when one goes on to quantize the system to obtain the spin foam model, the cotangent space variables get
quantized to dif\/ferential operators on the group.
Typically (for compact Lie groups), these geometric operators possess discrete spectra, and so the transparent classical
discrete geometry described by continuous metric variables gets replaced by the quantum geometry described by discrete
spin labels.
This corresponds to a~representation of the states and amplitudes of the model in terms of eigenstates of the geometric
operators, the spin representation~-- hence the name `spin' foams.
The quantum discreteness of geometric variables in spin foams, i.e., the use of quantum numbers as opposed to phase
space variables, although very useful to make contact with the canonical quantum theory, makes the amplitudes lose
a~direct contact with the classical discrete action and the classical discrete geometric variables.
The use of such classical discrete geometric variables, on the other hand, has been prevented until recently by their
non-commutative nature.

However, recently, a~new mathematical tool was introduced in the context of 3d quantum gravity, which became to be
called the `group Fourier transform'~\cite{BDOT,BGO,BO,BO2,BO3,DGO,FL,FM,JMN,OR}.
This is an $L^2$-isometric map from functions on a~Lie group to functions on the cotangent space equipped with
a~(generically) non-commutative $\star$-product structure.
In~\cite{GOR}, the transform was generalized to the `non-commutative Fourier transform' for all exponential Lie groups
by de\-ri\-ving it from the canonical symplectic structure of the cotangent bundle, and the non-commutative structure was
seen to arise from the deformation quantization of the algebra of geometric operators.
Accordingly, the non-commutative but continuous metric variables obtained through the non-commutative Fourier transform
correspond to the classical metric variables in the sense of deformation quantization.
Thus, it enables one to describe the quantum geometry of spin foam models and group f\/ield theory~\cite{BGO, BO} (and
Loop Quantum Gravity~\cite{BDOT,DGO}) by classical-like continuous metric variables.

The aim of this paper is to initiate the application of the above results in analysing the geometric properties of spin
foam models, in particular, in the classical limit ($\hbar \rightarrow 0$).
We will restrict our consideration to the 3d Ponzano--Regge model~\cite{Barrett08,Boulatov92, PonzanoRegge68} to have
a~better control over the formalism in this simpler case.
However, already for the Ponzano--Regge model we discover non-trivial properties of the metric representation related to
the non-commutative structure, which elucidate aspects of the use of non-commutative Fourier transform in the context of
spin foam models.
In particular, we f\/ind that in applying the stationary phase approximation one must account for the deformation
structure of the phase space in the variational calculus in order to recover the correct geometric constraints for the
metric variables in the classical limit of the phase space path integral.
Otherwise, the classical geometric interpretation of metric boundary data depends on the ambiguous choice of
quantization map for the algebra of geometric operators, which seems problematic.
Nevertheless, once the \emph{deformed} variational principle adapted to the non-commutative structure of the phase space
is employed, the non-commutative Fourier transform is seen to facilitate an unambiguous and straightforward asymptotic
analysis of the full partition function via a~\emph{non-commutative} stationary phase approximation.

In Section~\ref{sec:GFT} we will f\/irst outline the formalism of non-commutative Fourier transform, adapted
from~\cite{GOR} to the context of gravitational models.
In Section~\ref{sec:PR} we introduce the Ponzano--Regge model, seen as a~discretization of the continuum 3d BF theory.
In Section~\ref{sec:PRNC} we then apply the non-commutative Fourier transform to the Ponzano--Regge model to obtain
a~representation of the model in terms of non-commutative metric variables, and write down an explicit expression for
the quantum amplitude for f\/ixed metric boundary data on a~boundary with trivial topology.
In Section~\ref{sec:asymp} we further study the classical limit of the Ponzano--Regge amplitudes for f\/ixed metric
boundary data, and f\/ind that the results dif\/fer for dif\/ferent choices of non-commutative structures unless one
accounts for the deformation structure in the variational calculus.
When this is taken into account, the resulting semi-classical approximation coincides with what one expects from
a~discrete gravity path integral.
In particular, if one considers only the partial saddle point approximation obtained by varying the discrete connection
only, one f\/inds that the discrete path integral reduces to the one for 2nd order Regge action in terms of discrete
triad variables.
In Section~\ref{sec:6j} we consider in more detail the Ponzano--Regge amplitude with non-commutative metric boundary
data for a~single tetrahedron.
We recover the Regge action in the classical limit of the amplitude, and explain the connection of our calculation to
the previous studies of spin foam asymptotics in terms of coherent states.
Section~\ref{sec:cc} summarizes the obtained results and points to further research.

\section{Non-commutative Fourier transform for SU(2)}
\label{sec:GFT}

Our exposition of the non-commutative Fourier transform for ${\rm SU}(2)$ in this section follows~\cite{GOR},
adapted to the needs of quantum gravity models.
Originally, a~specif\/ic realization of the non-commutative Fourier transform formalism for the group ${\rm SO}(3)$ was
introduced in~\cite{FL} by Freidel \& Livine, and later expanded on by Freidel \& Majid~\cite{FM} and Joung, Mourad \&
Noui~\cite{JMN} to the case of ${\rm SU}(2)$.
(More abstract formulations of a~similar concept have appeared also in~\cite{MS, S}.) In our formalism this original
version of the transform corresponds to a~specif\/ic choice of a~quantization of the algebra of geometric operators,
which we will refer to as the Freidel--Livine--Majid quantization map, and treat it as one of the concrete examples we
give of the more general formulation in Subsection~\ref{subsec:examples}.\footnote{In addition, another realization of
the non-commutative Fourier transform for ${\rm SU}(2)$ relying on spinors was formulated by Dupuis, Girelli \& Livine
in~\cite{DGL}, but we will not consider it here.}

Let us consider the group ${\rm SU}(2)$, the Lie algebra $\textrm{Lie}({\rm SU}(2)) =: \mathfrak{su}(2)$ of ${\rm SU}(2)$, and the
associated cotangent bundle $\mathcal{T}^*{\rm SU}(2) \cong {\rm SU}(2) \times \mathfrak{su}(2)^*$.
As it is a~cotangent bundle, $\mathcal{T}^*{\rm SU}(2)$ carries a~canonical symplectic structure.
This is given by the Poisson brackets
\begin{gather}
\label{eq:poisson}
\{O,O'\} \equiv \frac{\partial O}{\partial X_i} \tilde{\mathcal{L}}_iO' - \tilde{\mathcal{L}}_iO \frac{\partial
O'}{\partial X_i} + \lambda \epsilon_{ij}^{\phantom{ij}k} \frac{\partial O}{\partial X_i} \frac{\partial O'}{\partial
X_j} X_k,
\end{gather}
where $O,O' \in C^\infty(\mathcal{T}^*{\rm SU}(2))$ are classical observables, and $\tilde{\mathcal{L}}_i:=
\lambda\mathcal{L}_i$ are dimensionful Lie derivatives on the group with respect to a~basis of right-invariant vector
f\/ields.
$\lambda\in\mathbb{R}_+$ is a~parameter with dimensions $[\frac{\hbar}{X}]$, which determines the physical scale
associated to the group manifold via the dimensionful Lie derivatives and the structure constants
$[\tilde{\mathcal{L}}_i,\tilde{\mathcal{L}}_j] = \lambda\epsilon_{ij}^{\phantom{ij}k}\tilde{\mathcal{L}}_k$.
$X_i$~are the Cartesian coordinates on $\mathfrak{su}(2)^*$.\footnote{Here it seems we are giving dimensions to
coordinates, which is usually a~bad idea in a~gravitational theory, to be considered below.
The point here is that the coordinates $X_i$ turn out to have a~geometric interpretation as discrete triad variables,
which is exactly what one would like to give dimensions to in general relativity.}

Let us now introduce coordinates $\zeta:{\rm SU}(2)\backslash \{-e\} \rightarrow \mathfrak{su}(2) \cong \mathbb{R}^3$ on the
dense subset ${\rm SU}(2)\backslash \{-e\} =: H \subset {\rm SU}(2)$, where $e \in {\rm SU}(2)$ is the identity element, which satisfy
\mbox{$\zeta(e) = 0$} and $\tilde{\mathcal{L}}_i\zeta^j(e) = \delta_i^j$.
The use of coordinates~$\zeta$ on~$H$ can be seen as a~sort of `one-point-decompactif\/i\-ca\-tion' of ${\rm SU}(2)$.
We then have for the Poisson brackets of the coordinates\footnote{Strictly speaking, the coordinates are not observables
of the classical system, but we may consider them def\/ined implicitly, since any observable may be parametrized in
terms of them, and they may be approximated arbitrarily closely by classical observables.}
\begin{gather*}
\{\zeta^i,\zeta^j\} = 0,
\qquad
\{X_i,\zeta^j\} = \tilde{\mathcal{L}}_i\zeta^j,
\qquad
\{X_i,X_j\} = \lambda\epsilon_{ij}^{\phantom{ij}k}X_k.
\end{gather*}
The Poisson brackets involving $\zeta^i$ are, of course, well-def\/ined only on~$H$.
We see that the commutators $\{X_i,\zeta^j\}$ of the chosen canonical variables are generically deformed due to the
curvature of the group manifold.
They coincide with the usual f\/lat commutation relations associated with Poisson-commuting coordinates only at the
identity.
Moreover, let us def\/ine \emph{the deformed addition} $\oplus_\zeta$ for these coordinates in the neighborhood of
identity as $\zeta(gh) =: \zeta(g) \oplus_\zeta \zeta(h)$.
It holds $\zeta(g) \oplus_\zeta \zeta(h) = \zeta(g) + \zeta(h) + \mathcal{O}(\lambda^0,|\ln(g)|,|\ln(h)|)$ for any
choice of~$\zeta$ complying with the above mentioned assumptions.
Indeed, the parametrization is chosen so that in the limit $\lambda \rightarrow 0$, while keeping the
coordinates~$\zeta$ f\/ixed, we ef\/fectively recover the f\/lat phase space $\mathcal{T}^*\mathbb{R}^3 = \mathbb{R}^3
\times \mathbb{R}^3 \cong \mathfrak{su}(2) \times \mathfrak{su}(2)^*$ from $\mathcal{T}^*{\rm SU}(2) = {\rm SU}(2) \times
\mathfrak{su}(2)^*$.
This follows because keeping~$\zeta$ f\/ixed implies a~simultaneous scaling of the class angles $|\ln(g)|$ of the group
elements.
Accordingly, the group ef\/fectively coincides with the tangent space $\mathfrak{su}(2)$ at the identity in this limit,
and~$\zeta$ become the Euclidean Poisson-commuting coordinates on $\mathfrak{su}(2) \cong \mathbb{R}^3$ for any initial
choice of~$\zeta$ satisfying the above assumptions.
Thus,~$\lambda$ can also be thought of as a~deformation parameter already at the level of the classical phase space.
For the above reasons, we will call the limit $\lambda \rightarrow 0$ the \emph{abelian} limit.

Let us then consider the quantization of the Poisson algebra given by the Poisson bracket~\eqref{eq:poisson}.
In particular, we consider the algebra $\mathfrak{H}$ generated by the operators $\hat{\zeta}^i$ and $\hat{X}_i$, modulo
the commutation relations
\begin{gather}
\label{eq:comrel}
[\hat{\zeta}^i,\hat{\zeta}^j] = 0,
\qquad
[\hat{X}_i,\hat{\zeta}^j] = i\hbar\widehat{\tilde{\mathcal{L}}_i\zeta^j},
\qquad
[\hat{X}_i,\hat{X}_j] = i\hbar\lambda\epsilon_{ij}^{\phantom{ij}k}\hat{X}_k.
\end{gather}
These relations follow from the symplectic structure of $\mathcal{T}^*{\rm SU}(2)$ in the usual way by imposing the relation
$[\mathfrak{Q}(O),\mathfrak{Q}(O')] \stackrel{!}{=} i\hbar\mathfrak{Q}(\{O,O'\})$ with the Poisson brackets of the
canonical variables, where by $\mathfrak{Q}: C^\infty(\mathcal{T}^*{\rm SU}(2)) \rightarrow \mathfrak{H}$ we denote the
quantization map specif\/ied by linearity, the ordering of operators, and $\mathfrak{Q}(\zeta^i) =: \hat{\zeta}^i$,
$\mathfrak{Q}(X_i) =: \hat{X}_i$.

We wish to represent the abstract algebra $\mathfrak{H}$ def\/ined by the commutation relations~\eqref{eq:comrel} as
operators acting on a~Hilbert space.
There exists the canonical representation in terms of smooth functions on $H \subset {\rm SU}(2)$ with the $L^2$-inner product
\begin{gather*}
\langle\psi|\psi'\rangle:= \frac{1}{\lambda^3} \int_{H} \dd g\, \overline{\psi(g)} \psi'(g),
\end{gather*}
where $\dd g$ is the normalized Haar measure, and the action of the canonical operators on is given~by
\begin{gather*}
\hat{\zeta}^i \psi \equiv \zeta^i \psi,
\qquad
\hat{X}_i \psi \equiv i\hbar\tilde{\mathcal{L}}_i\psi.
\end{gather*}
However, we would like to represent our original conf\/iguration space ${\rm SU}(2)$ rather than~$H$, and therefore we will
instead consider smooth functions on ${\rm SU}(2)$, whose restriction on~$H$ is clearly always in $C^\infty(H)$.
Since the coordinates are well-def\/ined only on $\mathcal{H} = {\rm SU}(2) \backslash \{-e\}$, the action of the coordinate
operators should then be understood only in a~weak sense: Even though strictly speaking the action $\hat{\zeta}^i \psi
\equiv \zeta^i \psi$ is not well-def\/ined for the whole of ${\rm SU}(2)$, the inner products
$\langle\psi|\hat{\zeta}^i|\psi'\rangle$ are, since we may write
\begin{gather*}
\langle\psi|\hat{\zeta}^i|\psi'\rangle = \frac{1}{\lambda^3} \int_{{\rm SU}(2)} \dd g \, \overline{\psi(g)} \zeta^i(g) \psi'(g)
\equiv \frac{1}{\lambda^3} \int_{H} \dd g\, \overline{\psi(g)} \zeta^i(g) \psi'(g)
\end{gather*}
for smooth~$\psi$, $\psi'$.
It is easy to verify that the commutation relations are represented correctly with this def\/inition of the action, and
the function space may be completed in the $L^2$-norm as usual.

However, there is also a~representation in terms of another function space, which is obtained through a~deformation
quantization procedure applied to the operator algebra corresponding to the other factor of the cotangent bundle,
$\mathfrak{su}(2)^*$
(see~\cite{GOR} for a~thorough exposition). Notice that the restriction of $\mathfrak{H}$ to the subalgebra generated~by
the operators $\hat{X}_i$ is isomorphic to a~completion of the universal enveloping algebra
$\overline{U(\mathfrak{su}(2))}$ of ${\rm SU}(2)$ due to its Lie algebra commutation relations.
A~$\star$-product for functions on $\mathfrak{su}(2)^*$ is uniquely specif\/ied by the restriction of the quantization
map $\mathfrak{Q}$ on the $\mathfrak{su}(2)^*$ part of the phase space via the relation $f \star f':=
\mathfrak{Q}^{-1}(\mathfrak{Q}(f)\mathfrak{Q}(f'))$, where $f,f' \in C^\infty(\mathfrak{su}(2)^*)$ and accordingly
$\mathfrak{Q}(f),\mathfrak{Q}(f') \in \overline{U(\mathfrak{su}(2))}$.
One may verify that the following action of the algebra on functions $\tilde{\psi} \in L^2_\star(\mathfrak{su}(2)^*)$
constitutes another representation of the algebra:
\begin{gather*}
\hat{\zeta}^i \tilde{\psi} \equiv -i\hbar\frac{\partial \tilde{\psi}}{\partial X_i},
\qquad
\hat{X}_i \tilde{\psi} \equiv X_i \star \tilde{\psi}.
\end{gather*}
The \emph{non-commutative Fourier transform} acts as an intertwiner between the canonical representation in terms of
square-integrable functions on ${\rm SU}(2)$ and the non-commutative dual space $L^2_\star(\mathfrak{su}(2)^*)$ of
square-integrable functions on $\mathfrak{su}(2)^*$ with respect to the $\star$-product.
It is given~by
\begin{gather*}
\tilde{\psi}(X) \equiv \int_{H} \frac{\dd g}{\lambda^3} \overline{E(g,X)} \psi(g) \in L^2_\star(\mathfrak{su}(2)^*),
\qquad
\psi \in L^2({\rm SU}(2)),
\end{gather*}
where the integral kernel
\begin{gather*}
E(g,X) \equiv e_\star^{\frac{i}{\hbar\lambda}k(g)\cdot X}:= \sum\limits_{n=0}^\infty \frac{1}{n!}
\left(\frac{i}{\hbar\lambda}\right)^n k^{i_1}(g) \cdots k^{i_n}(g) X_{i_1} \star \dots \star X_{i_n}
\end{gather*}
is the \emph{non-commutative plane wave}, and we denote $k(g):=-i\ln(g) \in \mathfrak{su}(2)$ taken in the principal
branch of the logarithm.
The inverse transform reads
\begin{gather*}
\psi(g) = \int_{\mathfrak{su}(2)^*} \frac{\dd X}{(2\pi\hbar)^3} E(g,X) \star \tilde{\psi}(X)\in L^2({\rm SU}(2)),
\qquad
\tilde{\psi} \in L_\star^2(\mathfrak{su}(2)^*),
\end{gather*}
where $\dd X:= \dd X_1\dd X_2\dd X_3$ denotes the Lebesgue measure on the Lie algebra dual
$\mathfrak{su}(2)^*\cong\mathbb{R}^3$.

Let us list some important properties of the non-commutative plane waves that we will use in the following:
\begin{gather}
E(g,X) = e_\star^{\frac{i}{\hbar\lambda}k(g)\cdot X} \equiv c(g)e^{\frac{i}{\hbar}\zeta(g)\cdot X},
\qquad
\text{where}
\qquad
c(g):= E(g,0),
\label{eq:Eform}
\\
\overline{E(g,X)} = E\big(g^{-1},X\big) = E(g,-X),
\nonumber
\\
E(\ad_hg,X) = E(g,\Ad_h^{-1}X),
\label{eq:Ead}
\\
E(gh,X) = E(g,X) \star E(h,X),
\label{eq:Eprod}
\\
\int_{\mathfrak{su}(2)^*}\frac{\dd X}{(2\pi\hbar\lambda)^3} E(g,X) = \delta(g),
\label{eq:Edelta}
\\
\tilde{\psi}(X) \star E(g,X) = E(g,X) \star \tilde{\psi}(\Ad_g X),
\label{eq:Eperm}
\end{gather}
where $\ad_hg:= hgh^{-1}$ and $\Ad_hX:= hXh^{-1}$.
Notice that from~\eqref{eq:Eform} and~\eqref{eq:Ead} it follows that $c(\ad_hg) = c(g)$ and $\zeta(\ad_hg) =
h\zeta(g)h^{-1} =: \Ad_h\zeta(g)$.
In addition, we f\/ind that the function
\begin{gather*}
\delta_\star(X,Y):= \int_H \frac{\dd g}{(2\pi\hbar\lambda)^3} \overline{E(g,X)} E(g,Y)
\end{gather*}
acts as the delta distribution with respect to the $\star$-product, namely,
\begin{gather*}
\int_{\mathfrak{su}(2)^*}\dd Y\, \delta_\star(X,Y) \star \tilde{\psi}(Y) = \tilde{\psi}(X) = \int_{\mathfrak{su}(2)^*}\dd Y
\, \tilde{\psi}(Y) \star \delta_\star(X,Y).
\end{gather*}
More generally, $\delta_\star$ is the integral kernel of the projection
\begin{gather*}
\mathcal{P}(\tilde{\psi})(X):= \int_{\mathfrak{su}(2)^*}\dd Y\, \delta_\star(X,Y) \star \tilde{\psi}(Y)
\end{gather*}
onto the image $L_\star^2(\mathfrak{su}(2)^*)$ of the non-commutative Fourier transform.
In the following, we will also occasionally slightly abuse notation by writing
\begin{gather*}
\delta_\star\left(\sum\limits_i X_i\right):= \int_H \frac{\dd g}{(2\pi\hbar\lambda)^3} \prod\limits_i E(g,X_i)
\end{gather*}
for convenience, although this is not a~function of the linear sum $\sum\limits_i X_i$ if $c(g)\neq 1$ for some $g\in
H$.

Finally, we wish to emphasize that the non-commutative coordinate variables of the dual representation are unambiguously
identif\/ied with the corresponding classical conjugate momenta to the group elements via deformation quantization.
This follows directly from the construction.
Indeed, it is a~key advantage of the above construction for the non-commutative representation that it retains a~direct
relation to the classical phase space quantities, thus helping to make the interpretation of the quantum expressions
more intuitive and straightforward, especially in the semi-classical regime.
Our primary goal in this paper is exactly to use this clear-cut relation to our benef\/it in analysing and interpreting
in discrete geometric terms the leading order semi-classical behavior of the Ponzano--Regge model.

\section{3d BF theory and the Ponzano--Regge model}
\label{sec:PR}
The Ponzano--Regge model can be understood as a~discretization of 3-dimensional Riemannian BF theory.
In this section, we will brief\/ly review how it can be derived from the continuum BF theory, while keeping track of the
dimensionful physical constants determining the various asymptotic limits of the theory.

Let $\mathcal{M}$ be a~3-dimensional base manifold to a~frame bundle with the structure group ${\rm SU}(2)$.
Then the partition function of 3d BF theory on $\mathcal{M}$ is given~by
\begin{gather}
\label{eq:BFaction}
\mathcal{Z}_{\rm BF}^{\mathcal{M}} = \int \mathcal{D} E \mathcal{D} \omega \exp\left(\frac{i}{2\hbar\kappa}
\int_{\mathcal{M}} \tr \big(E \wedge F(\omega) \big) \right),
\end{gather}
where~$E$ is an $\mathfrak{su}(2)^*$-valued triad 1-form on $\mathcal{M}$, $F(\omega)$ is the $\mathfrak{su}(2)$-valued
curvature 2-form associated to the connection 1-form~$\omega$, and the trace is taken in the fundamental
spin-$\frac{1}{2}$ representation of ${\rm SU}(2)$.
$\hbar$ is the reduced Planck constant and~$\kappa$ is a~constant with dimensions of inverse momentum.
The connection with Riemannian gravity in three spacetime dimensions gives $\kappa:= 8\pi G$, where~$G$ is the
gravitational constant.
Since the triad 1-form~$E$ has dimensions of length and the curvature 2-form~$F$ is dimensionless, the exponential is
rendered dimensionless by dividing with $\hbar\kappa \equiv 8\pi l_p$, $l_p \equiv \hbar G$ being the Planck length in
three dimensions.
Integrating over the triad f\/ield in~\eqref{eq:BFaction}, we get heuristically
\begin{gather}
\label{eq:BFdelta}
\mathcal{Z}_{\rm BF}^{\mathcal{M}} \propto \int \mathcal{D} \omega\, \delta\big(F(\omega)\big),
\end{gather}
so we see that the BF partition function is (at least nominally) nothing but the volume of the moduli space of f\/lat
connections on $\mathcal{M}$.\footnote{The volume of a~moduli space can be def\/ined via its natural symplectic
structure, and in some 2-dimensional cases has been rigorously related to a~QFT partition function,
see~\cite{Goldman1984,Sengupta2003,Witten1991}.} Generically, this is of course divergent, which (among other things)
motivates us to consider discretizations of the theory.
However, since BF theory is purely topological, that is, it does not depend on the metric structure of the base
manifold, such a~discretization should not af\/fect its essential properties.

\begin{figure}[t]
\centering
\includegraphics{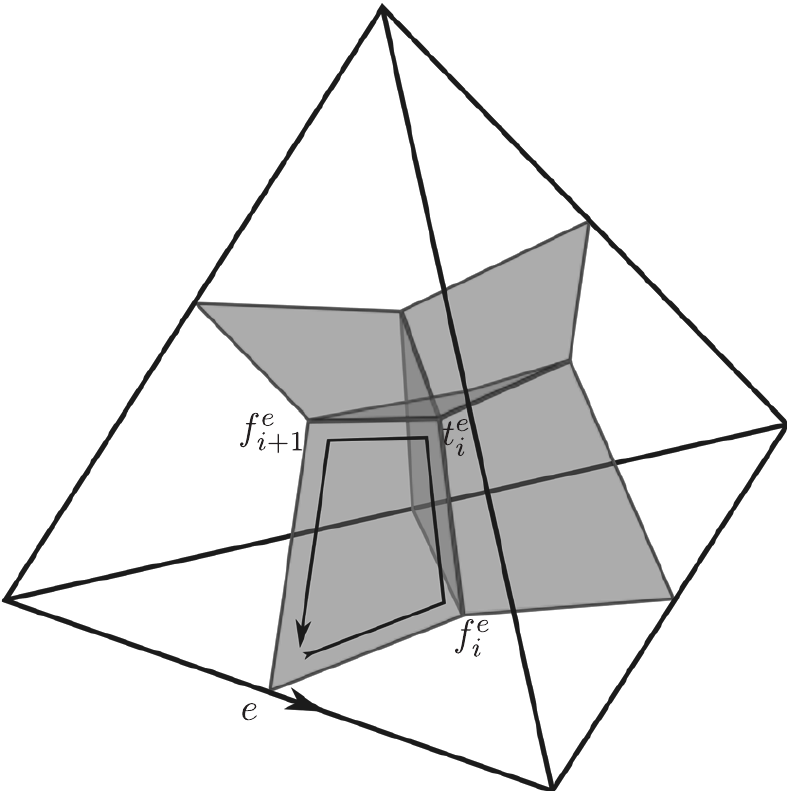}
 \caption{The subdivision of tetrahedra in~$\Delta$ into a~f\/iner cellular complex~$\Gamma$.
Here,~$e$ labels an edge of the triangulation, $f^e_i$ label (the centers of) the triangles incident to the edge~$e$,
and $t^e_i$ label (the centers of) the tetrahedra incident to~$e$.
The index~$i$ runs from 0 to $n_e-1$, $n_e$ being the number of triangles incident to~$e$.}\label{fig:tetrasubdiv}
\end{figure}

Now, to discretize the continuum BF theory, we f\/irst choose a~triangulation~$\Delta$ of the mani\-fold~$\mathcal{M}$,
that is, a~(homogeneous) simplicial complex homotopic to~$\mathcal{M}$.
The dual complex~$\Delta^*$ of~$\Delta$ is obtained by replacing each~$d$-simplex in~$\Delta$ by a~$(3-d)$-simplex and
retaining the connective relations between simplices.
Then, the homotopy between~$\Delta$ and $\mathcal{M}$ allows us to think of~$\Delta$, and thus $\Delta^*$, as embedded
in $\mathcal{M}$.
We further form a~f\/iner cellular complex~$\Gamma$ by diving the tetrahedra in~$\Delta$ along the faces of $\Delta^*$.
In particular,~$\Gamma$ then consists of tetrahedra $t \in \Delta$, with vertices $t^* \in \Delta^*$ at their centers,
each subdivided into four cubic cells.
Moreover, for each tetrahedron $t \in \Delta$, there are edges $tf \in \Gamma$, which correspond to half-edges of $f^*
\in \Delta^*$, going from the centers of the triangles $f \in \Delta$ bounding the tetrahedron to the center of the
tetrahedron~$t$.
Also, for each triangle $f \in \Delta$, there are edges $ef \in \Gamma$, which go from the center of the triangle $f \in
\Delta$ to the centers of the edges $e \in \Delta$ bounding the triangle~$f$.
See Fig.~\ref{fig:tetrasubdiv} for an illustration of the subdivision of a~single tetrahedron in~$\Delta$.

To obtain the discretized connection variables associated to the triangulation~$\Delta$, we integrate the connection
along the edges $tf \in \Gamma$ and $ef \in \Gamma$ as
\begin{gather*}
g_{tf}:= \mathcal{P} e^{i \int_{tf} \omega} \in {\rm SU}(2)
\qquad
\text{and}
\qquad
g_{ef}:= \mathcal{P} e^{i \int_{ef} \omega} \in {\rm SU}(2),
\end{gather*}
where $\mathcal{P}$ denotes the path-ordered exponential.
Thus, they are the Wilson line variables of the connection~$\omega$ associated to the edges or, equivalently, the
parallel transports from the initial to the f\/inal points of the edges with respect to~$\omega$.
We assume the triangulation~$\Delta$ to be piece-wise f\/lat, and associate frames to all simplices of~$\Delta$.
We then interpret $g_{tf}$ as the group element relating the frame of $t \in \Delta$ to the frame of $f \in \Delta$, and
similarly $g_{ef}$ as the group element relating the frame of $f \in \Delta$ to the frame of $e \in \Delta$.
Furthermore, we integrate the triad f\/ield along the edges $e \in \Delta$ as
\begin{gather}
\label{eq:Xdef}
X_e:= \int_{e} \Ad_{G_e} E \in \mathfrak{su}(2)^*.
\end{gather}
Here, $G_e$ denotes the ${\rm SU}(2)$-valued function on the edge~$e$ that parallel transports via adjoint action the
pointwise values of~$E$ along~$e$ to a~f\/ixed base point at the center of~$e$.
An orientation for the edge~$e$ may be chosen arbitrarily.
$X_e$ is interpreted as the vector representing the magnitude and the direction of the edge~$e$ in the frame associated
to the edge~$e$ itself.

In the case that~$\Delta$ has no boundary, a~discrete version of the BF partition function~\eqref{eq:BFdelta}, the
Ponzano--Regge partition function, may be written as
\begin{gather*}
\mathcal{Z}_{\rm PR}^\Delta = \int \bigg[\prod\limits_{tf} \dd g_{tf}\bigg] \prod\limits_{e \in \Delta}
\delta(H_{e^*}(g_{tf})),
\end{gather*}
where $H_{e^*}(g_{tf})\in {\rm SU}(2)$ are holonomies around the dual faces $e^*\in\Delta^*$ obtained as products of~$g_{tf}$,
$f^*\in \partial e^*$, and~$\dd g_{tf}$ is again the Haar measure on~${\rm SU}(2)$.
Mimicking the continuum partition function of BF theory, the Ponzano--Regge partition function is thus an integral over
the f\/lat discrete connections, the delta functions $\delta(H_{e^*}(g_{tf}))$ constraining holonomies around all dual
faces to be trivial.

Now, we can apply the non-commutative Fourier transform to expand the delta functions in terms of non-commutative plane
waves by equation~\eqref{eq:Edelta}.
This yields
\begin{gather}
\mathcal{Z}_{\rm PR}^\Delta = \int \bigg[\prod\limits_{tf} \dd g_{tf}\bigg] \bigg[\prod\limits_{e} \frac{\dd
X_e}{(2\pi\hbar\lambda)^3}\bigg] \bigg[\prod\limits_{e \in \Delta} c(H_{e^*}(g_{tf})) \bigg] \exp\bigg\{\frac{i}{\hbar}
\sum\limits_{e\in\Delta} X_e \cdot \zeta(H_{e^*}(g_{tf})) \bigg\}.\!\!\!
\label{discreteBFpathint}
\end{gather}
Comparing with~\eqref{eq:BFaction}, this expression has a~straightforward interpretation as a~discretization of the
f\/irst order path integral of the continuum BF theory.
We can clearly identify the discretized triad variables $X_e$ in~\eqref{eq:Xdef} with the non-commutative metric
variables def\/ined via the non-commutative Fourier transform.
We also see that, from the point of view of discretization, the form of the plane waves and thus the choice for the
quantization map is directly related to the choice of the precise form for the discretized action and the path integral
measure.
In particular, the coordinate function $\zeta: {\rm SU}(2) \rightarrow \mathfrak{su}(2)$ and the prefactor
$c:{\rm SU}(2)\rightarrow\mathbb{C}$ of the non-commutative plane wave are dictated by the choice of the quantization map, and
the coordinates specify the discretization prescription for the curvature 2-form $F(\omega)$.
Similar interplay between $\star$-product quantization and discretization is well-known in the case of the f\/irst order
phase space path integral formulation of ordinary quantum mechanics~\cite{CD}.
Moreover, on dimensional grounds, we must identify $\lambda\equiv\kappa=8\pi G$, so that the coordinates~$\zeta$ have
the dimensions of $\frac{1}{\kappa}F(\omega)$.
Therefore, the abelian limit of the non-commutative structure of the phase space corresponds in this case also
physically to the no-gravity limit $G\rightarrow 0$.
We will denote this classical deformation parameter associated with the non-commutative structure of the phase space
collectively by~$\kappa$ in the following.

\section{Non-commutative metric representation
\\
of the Ponzano--Regge model}
\label{sec:PRNC}

If the triangulated manifold~$\Delta$ has a~non-trivial boundary, we may assign connection data on the boundary~by
f\/ixing the group elements $g_{ef}$ associated to the boundary triangles $f \in \partial\Delta$.
Then, the (non-normalized) Ponzano--Regge amplitude for the boundary can be written as
\begin{gather}
\label{eq:boundamp}
\mathcal{A}_{\rm PR}(g_{ef}|f\in\partial\Delta) = \int \bigg[\prod\limits_{tf} \dd g_{tf}\bigg]
\bigg[\prod\limits_{\substack{ef
\\
f\notin\partial\Delta}} \dd g_{ef}\bigg] \prod\limits_{e \in \Delta} \prod\limits_{i=0}^{n_e-1} \delta\big(g_{ef_{i+1}^e}
g_{t_i^ef_{i+1}^e}^{-1} g_{t_i^ef_i^e} g_{ef_i^e}^{-1} \big).
\end{gather}
The delta functions are over the holonomies around the wedges of the triangulation pictured in grey in
Fig.~\ref{fig:tetrasubdiv}.
For this purpose, the tetrahedra $t^e_i$ and the triangles $f^e_i$ sharing the edge~$e$ are labelled by an index
$i=0,\ldots,n_e-1$ in a~right-handed fashion with respect to the orientation of the edge~$e$ and with the
identif\/ication $f_{n_e}\equiv f_0$, as in Fig.~\ref{fig:tetrasubdiv}.

\begin{figure}[t]
\centering \includegraphics{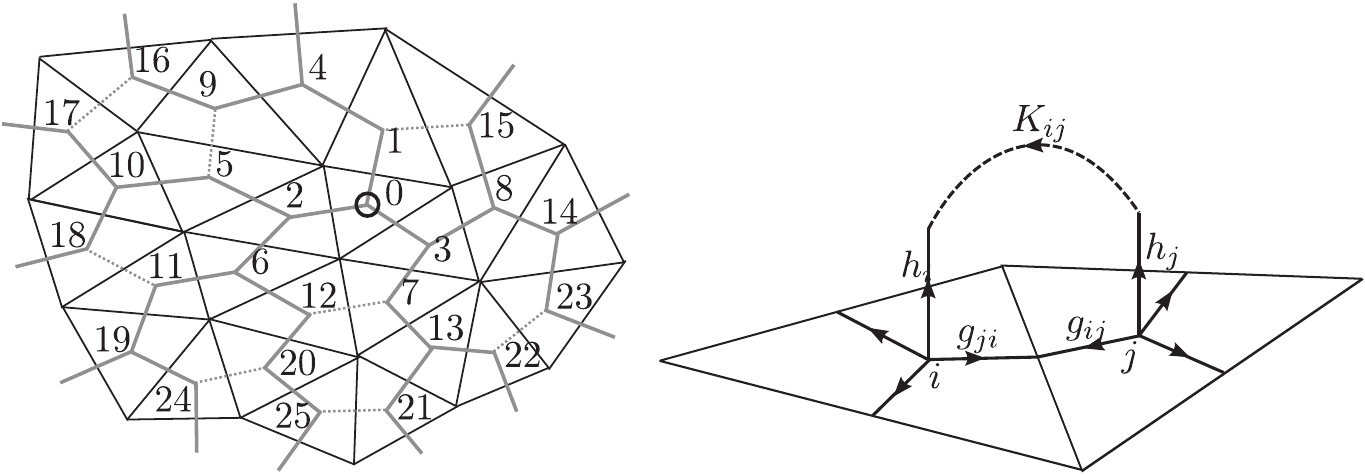}
\caption{On the left: A~portion of a~rooted labelled spanning tree of the dual graph of a~boundary triangulation (solid grey
edges).
On the right: Boundary triangles $f_i,f_j \in \partial\Delta$ and the associated group elements.}\label{fig:treeholos}
\end{figure}

Let us introduce some simplifying notation.
We will choose an arbitrary spanning tree of the dual graph to the boundary triangulation, pick an arbitrary root vertex
for the tree, and label the boundary triangles $f_i \in \partial\Delta$ by $i \in \mathbb{N}_0$ in a~compatible way with
respect to the partial ordering induced by the tree, so that the root has the label~$0$
(see Fig.~\ref{fig:treeholos}). Moreover, we denote the set of ordered pairs of labels associated to neighboring
boundary triangles by $\mathcal{N}$, and label the group elements associated to the pair of neighboring boundary
triangles $(i,j)\in\mathcal{N}$ as illustrated in Fig.~\ref{fig:treeholos}.
The group elements $g_{tf}$, $f \notin \partial\Delta$, will be denoted by a~collective label $h_l$.
As we integrate over $g_{ef}$ for $f \notin \partial\Delta$ in~\eqref{eq:boundamp}, we obtain
\begin{gather}
\label{eq:BFdeltas}
\mathcal{A}_{\rm PR}(g_{ij}) = \int \bigg[\prod\limits_{l} \dd h_{l}\bigg] \bigg[\prod\limits_{e \notin \partial\Delta}
\delta(H_{e^*}(h_l))\bigg] \bigg[\prod\limits_{\substack{(i,j)\in\mathcal{N}
\\
i<j}} \delta\big(g_{ij} h_{j}^{-1} K_{ji}(h_l) h_{i} g_{ji}^{-1}\big) \bigg].
\end{gather}
Here $h_i$ is the group element associated to the dual half-edge going from the boundary triangle~$i$ to the center of
the tetrahedron with triangle~$i$ on its boundary, and $K_{ij}(h_l)$ is the holonomy along the bulk dual edges from the
center of the tetrahedron with triangle~$j$ to the center of the tetrahedron with triangle~$i$
(see Fig.~\ref{fig:treeholos} for illustration). There is a~one-to-one correspondence between the pairs $(i,j)$ of
neighboring boundary triangles and faces of the dual 2-complex touching the boundary.
Notice that we have chosen here as the base point of each holonomy the boundary dual vertex with a~smaller label.
By expanding the delta distributions in~\eqref{eq:BFdeltas} with boundary group variables into non-commutative plane
waves, we get
\begin{gather}
\mathcal{A}_{\rm PR}(g_{ij})=\int \bigg[\prod\limits_l\dd h_l\bigg] \bigg[\prod\limits_{e \notin
\partial\Delta}\delta(H_{e^*}(h_l)) \bigg]
\nonumber
\\
\phantom{\mathcal{A}_{\rm PR}(g_{ij})=} \times\Bigg[\prod\limits_{\substack{(i,j)\in\mathcal{N}
\\
i<j}} \int\frac{\dd Y_{ji}}{(2\pi\hbar\kappa)^3}E\big(g_{ij}h_{j}^{-1} K_{ji}(h_l) h_{i} g_{ji}^{-1},Y_{ji}\big) \Bigg].
\label{eq:deltaexp}
\end{gather}

To obtain the expression for metric boundary data, we employ the non-commutative Fourier transform,
\begin{gather}
\label{eq:metricreps}
\tilde{\mathcal{A}}_{\rm PR}(X_{ij}) = \int \bigg[\prod\limits_{(i,j)\in\mathcal{N}} \frac{\dd g_{ij}}{\kappa^3}\bigg]
\mathcal{A}_{\rm PR}(g_{ij}) \prod\limits_{(i,j)\in\mathcal{N}} E\big(g_{ij}^{-1},X_{ij}\big).
\end{gather}
Here the variable $X_{ij}$ is understood geometrically as the edge vector shared by the triangles~$i$,~$j$ as seen from
the frame of reference of the triangle~$j$.
We note that the exact functional form of the amplitude, as that of the non-commutative plane wave, depends on the
particular choice of a~quantization map.
From~\eqref{eq:deltaexp} and~\eqref{eq:metricreps} the amplitude for metric boundary data is obtained by expanding the
delta functions as
\begin{gather}
\tilde{\mathcal{A}}_{\rm PR}(X_{ij})=\int \bigg[\prod\limits_{(i,j)\in\mathcal{N}} \frac{\dd g_{ij}}{\kappa^3}\bigg]
\left[\frac{\dd Y_{ji}}{(2\pi\hbar\kappa)^3}\right] \bigg[\prod\limits_l\dd h_l\bigg] \left[\frac{\dd
Y_{e}}{(2\pi\hbar\kappa)^3}\right] \bigg[\prod\limits_{e \notin \partial\Delta} E(H_{e^*}(h_l),Y_e) \bigg]
\nonumber
\\
\phantom{\tilde{\mathcal{A}}_{\rm PR}(X_{ij})=} \times \bigg[\prod\limits_{\substack{(i,j)\in\mathcal{N}
\\
i<j}} E\big(g_{ij} h_{j}^{-1} K_{ji}(h_l) h_{i} g_{ji}^{-1},Y_{ji}\big) \bigg] \bigg[\prod\limits_{(i,j)\in\mathcal{N}}
E\big(g_{ij}^{-1},X_{ij}\big) \bigg].
\label{eq:NCpwmetricreps}
\end{gather}
We emphasize that here $X_{ij}$'s are the f\/ixed boundary edge vectors, while $Y_{ji}$'s are auxiliary boundary edge
vectors, which are the Lagrange multipliers imposing the triviality of holonomies around dual faces touching the
boundary.
We will see that the two are identif\/ied (up to orientations and parallel transports) in the classical limit.
Importantly, equation~\eqref{eq:NCpwmetricreps} is nothing else than the simplicial path integral for a~complex with
boundary, and a~f\/ixed discrete metric on this boundary represented by~$X_{ij}$'s.
This can be seen by writing the explicit form of the non-commutative plane waves, thus obtaining a~formula
like~\eqref{discreteBFpathint}, augmented by boundary terms.
We will use this expression in the next section to study the semi-classical limit.

\subsection*{Exact amplitudes for metric boundary data on a~sphere}

By integrating over all $Y_e$ and using the property~\eqref{eq:Eprod} for the non-commutative plane waves, we may
write~\eqref{eq:NCpwmetricreps} as
\begin{gather}
\tilde{\mathcal{A}}_{\rm PR}(X_{ij}) \propto \int \bigg[\prod\limits_{(i,j)\in\mathcal{N}} \frac{\dd g_{ij}}{\kappa^3}\bigg]
\left[\frac{\dd Y_{ji}}{(2\pi\hbar\kappa)^3}\right] \bigg[\prod\limits_l\dd h_l\bigg] \bigg[\prod\limits_{e \notin
\partial\Delta} \delta(H_{e^*}(h_l)) \bigg]\label{eq:alice}
\\
\times \bigg[\prod\limits_{\substack{(i,j)\in\mathcal{N}
\\
i<j}} \big(E(g_{ij},Y_{ji})E\big(g_{ij}^{-1},X_{ij}\big)\big) \star E(h_{j}^{-1} K_{ji}(h_l) h_{i},Y_{ji}) \star
\big(E\big(g_{ji}^{-1},Y_{ji}\big) E\big(g_{ji}^{-1},X_{ji}\big)\big) \bigg],\nonumber
\end{gather}
where the $\star$-product acts on $Y_{ji}$.
For simplicity, we often do not include explicitly the f\/inite proportionality constants in front of amplitudes,
because they are immaterial for our results, and will eventually be cancelled by normalization.
Further integrating in~\eqref{eq:alice} over all $g_{ij}$ gives
\begin{gather*}
\tilde{\mathcal{A}}_{\rm PR}(X_{ij}) \propto \int \left[\frac{\dd Y_{ji}}{(2\pi\hbar\kappa)^3}\right] \left[\prod\limits_l\dd
h_l\right] \left[\prod\limits_{e \notin \partial\Delta} \delta(H_{e^*}(h_l)) \right]
\\
\phantom{\tilde{\mathcal{A}}_{\rm PR}(X_{ij}) \propto}{}
\times \left[\prod\limits_{\substack{(i,j)\in\mathcal{N}
\\
i<j}} \delta_\star(Y_{ji},X_{ij}) \star E\big(h_{j}^{-1} K_{ji}(h_l) h_{i},Y_{ji}\big) \star \delta_\star(Y_{ji},-X_{ji})
\right],
\end{gather*}
where now the $\delta_\star$-functions impose the identif\/ications of boundary edge vector variables, up to parallel
transport.
Indeed, the non-commutative plane wave takes care of the parallel transport between the frames of $X_{ij}$ and $X_{ji}$,
as we may easily observe using the property~\eqref{eq:Eperm} of the plane wave as we permute the f\/irst
$\delta_\star$-function with the plane wave to obtain
\begin{gather*}
\tilde{\mathcal{A}}_{\rm PR}(X_{ij}) \propto \int \left[\frac{\dd Y_{ji}}{(2\pi\hbar\kappa)^3}\right] \bigg[\prod\limits_l\dd
h_l\bigg] \bigg[\prod\limits_{e \notin \partial\Delta} \delta(H_{e^*}(h_l)) \bigg]
\nonumber
\\
\hphantom{\tilde{\mathcal{A}}_{\rm PR}(X_{ij}) \propto}{}
\times \bigg[\prod\limits_{\substack{(i,j)\in\mathcal{N}
\\
i<j}} E\big(h_{j}^{-1} K_{ji}(h_l) h_{i},Y_{ji}\big) \star \delta_\star\big(\Ad_{h_{j}^{-1} K_{ji}(h_l) h_{i}} Y_{ji},X_{ij}\big) \star
\delta_\star(Y_{ji},-X_{ji}) \bigg].
\end{gather*}
We may further integrate over all $Y_{ji}$ to get
\begin{gather*}
\tilde{\mathcal{A}}_{\rm PR}(X_{ij}) \propto \int \bigg[\prod\limits_l\dd h_l\bigg] \bigg[\prod\limits_{e \notin
\partial\Delta} \delta(H_{e^*}(h_l)) \bigg]
\\
\hphantom{\tilde{\mathcal{A}}_{\rm PR}(X_{ij}) \propto}{}
\times \bigg[\prod\limits_{\substack{(i,j)\in\mathcal{N}
\\
i<j}} E\big(h_{i}^{-1} K_{ij}(h_l) h_{j},X_{ji}\big) \star \delta_\star\big(\Ad_{h_{j}^{-1} K_{ji}(h_l) h_{i}} X_{ji},-X_{ij}\big)
\bigg].
\end{gather*}
We see that the edge vectors $X_{ij}$, $X_{ji}$ corresponding to the same edge (with opposite orientations) in
dif\/ferent frames of reference are identif\/ied up to a~parallel transport by $h_{j}^{-1} K_{ji}(h_l) h_{i}$ through
the non-commutative delta distributions $\delta_\star(\Ad_{h_{j}^{-1} K_{ji}(h_l) h_{i}} X_{ji},-X_{ij})$.

We wish to further integrate over the variables $h_i$.
To this aim, we employ the change of variables $X_{ji} \mapsto \Ad_{h_{i}^{-1} K_{ij}(h_l) h_{j}} X_{ji}$, i.e., we
parallel transport the variables~$X_{ji}$ to the frames of~$X_{ij}$ to get a~simple identif\/ication of the boundary
variables, and move all $h_i$-dependence to the plane waves.
We thus get
\begin{gather*}
\begin{split}
& \tilde{\mathcal{A}}_{\rm PR}(X_{ij}) \propto \int \bigg[\prod\limits_l\dd h_l\bigg] \bigg[\prod\limits_{e \notin
\partial\Delta} \delta(H_{e^*}(h_l)) \bigg]
\\
& \hphantom{\tilde{\mathcal{A}}_{\rm PR}(X_{ij}) \propto}{}
\times \bigg[\prod\limits_{\substack{(i,j)\in\mathcal{N}
\\
i<j}} E\big(h_{i}^{-1}K_{ij}(h_l)h_{j},X_{ji}\big) \bigg] \star \bigg[\prod\limits_{\substack{(i,j)\in\mathcal{N}
\\
i<j}} \delta_{\star}(X_{ij},-X_{ji}) \bigg],
\end{split}
\end{gather*}
Note that for every vertex~$i$ there is a~unique path via the edges $(j_{n-1},j_n)_{n=1,\ldots,l}$, s.t.~$j_{0}=0$,
$j_l=i$, from the root to the vertex~$i$ along the spanning tree.
Now, by making the changes of variables
\begin{gather*}
h_i \mapsto \left[\mathop{\overleftarrow{\prod}}\limits_{n=0}^{l} K_{j_{n-1}j_{n}}^{-1}(h_l) \right] h_i,
\end{gather*}
where by $\overleftarrow{\prod}$ we denote an ordered product of group elements such that the product index increases
from right to left, we obtain
\begin{gather*}
\tilde{\mathcal{A}}_{\rm PR}(X_{ij}) \propto \int \bigg[\prod\limits_l\dd h_l\bigg] \bigg[\prod\limits_{e \notin
\partial\Delta} \delta(H_{e^*}(h_l)) \bigg] \bigg[\prod\limits_{\substack{(i,j) \in \textrm{tree}
\\
i<j}} E(h_{i}^{-1}h_{j},X_{ji}) \bigg]
\\
\hphantom{\tilde{\mathcal{A}}_{\rm PR}(X_{ij}) \propto}{}
\times \bigg[\prod\limits_{\substack{(i,j) \notin \textrm{tree}
\\
i<j}} E(h_{i}^{-1}L_{ij}(h_l)h_{j},X_{ji}) \bigg] \star \bigg[\prod\limits_{\substack{(i,j)\in\mathcal{N}
\\
i<j}} \delta_{\star}(X_{ij}, -X_{ji}) \bigg]
\\
\hphantom{\tilde{\mathcal{A}}_{\rm PR}(X_{ij}) }{}
=\int \bigg[\prod\limits_l\dd h_l\bigg] \bigg[\prod\limits_{e \notin\partial\Delta} \delta(H_{e^*}(h_l)) \bigg]
\\
\hphantom{\tilde{\mathcal{A}}_{\rm PR}(X_{ij}) \propto}{}
\times \Bigg[\rsprod{i}{} \bigg(E(h_{i},\sum\limits_{j} \epsilon_{ij} X_{ji})  \star \!\prod\limits_{\substack{j
\\
(i,j) \notin \textrm{tree}}} E(L_{ij}(h_l),X_{ji}) \bigg) \Bigg] \!\star
\!\bigg[\prod\limits_{\substack{(i,j)\in\mathcal{N}
\\
i<j}} \! \delta_{\star}(X_{ij},-X_{ji}) \bigg].
\end{gather*}
Here, $\epsilon_{ij}:= \sgn(i-j)A_{ij}$, where $A_{ij}$ is the adjacency matrix of the dual graph of the boundary
triangulation.
Moreover, $L_{ij}(h_l) \equiv G_{ij}^{-1}(h_l)H_{ij}(h_l)G_{ij}(h_l)$, where $H_{ij}(h_l)$ is the product of
$K_{kl}(h_l)$'s around the unique cycle of the boundary dual graph formed by adding the edge $(i,j)$ to the spanning
tree, and $G_{ij}(h_l)$ is the product of $K_{kl}(h_l)$'s along the unique path from the root of the spanning tree to
the cycle.
The cycles formed from the spanning tree of a~graph by adding single edges span the loop space of the graph.
Thus, $H_{ij}(h_l)$ are trivial for a~trivial boundary topology, if the product of $K_{kl}(h_l)$'s around all boundary
vertices are trivial.
On the other hand, the product of $K_{kl}(h_l)$'s around a~boundary vertex is constrained to be trivial by the
f\/latness constraints for the bulk holonomies only if the neighborhood of the vertex is a~half-ball, since only in this
case is the loop around the vertex contractible along the faces of the 2-complex.
Thus, given that the neighborhoods of all boundary vertices have trivial topology, the f\/latness constraints impose
$L_{ij}(h_l)$ to be trivial, if the boundary has a~trivial topology, i.e., $\partial\Delta \cong S^2$.
Accordingly, we have
\begin{gather*}
\tilde{\mathcal{A}}_{\rm PR}(X_{ij}) \propto \int \bigg[\prod\limits_l\dd h_l\bigg] \bigg[\prod\limits_{e \notin
\partial\Delta} \delta(H_{e^*}(h_l)) \bigg]
\\
\hphantom{\tilde{\mathcal{A}}_{\rm PR}(X_{ij}) \propto}{}
\times\left[\rsprod{i}{} E(h_{i},\epsilon_{ij} X_{ji}) \right] \star \bigg[\prod\limits_{\substack{(i,j)\in\mathcal{N}
\\
i<j}} \delta_{\star}(X_{ij},-X_{ji}) \bigg],
\end{gather*}
where we used the notation $\overstar\overrightarrow{\prod}$ for the ordered star product of plane waves.
Integrating over $h_i$ then yields the closure constraints for the boundary triangles, and we end up with
\begin{gather}
\label{eq:metricBF}
\tilde{\mathcal{A}}_{\rm PR}(X_{ij}) \propto [\delta(0)]^{d} \left[\rsprod{i}{} \delta_{\star}\Big(\sum\limits_j \epsilon_{ij}
X_{ji}\Big) \right] \star \bigg[\prod\limits_{\substack{(i,j)\in\mathcal{N}
\\
i<j}} \delta_{\star}(X_{ij},-X_{ji}) \bigg],
\end{gather}
where the sum is over vertices~$j$ connected to the vertex~$i$, and~$d$ is the degree of divergence arising from the
redundant delta distributions over the dual faces $e^* \in \Delta^*$, $e \notin \partial\Delta$.

It is clear that in the abelian limit $\kappa\rightarrow 0$, where the $\star$-product coincides with the pointwise
product and $\delta_\star \rightarrow \delta$, the above amplitude imposes closure and identif\/ication of the edge
vectors.
However, the case of the classical limit $\hbar\rightarrow 0$ is more subtle: The whole notion of a~non-commutative
Fourier transform breaks down in this limit, since the non-commutative plane wave becomes ill-def\/ined, having no
well-def\/ined limit.
We will see in the following the ef\/fects of these complications and how to take them into account in studying the
classical limit.

\section{Semi-classical analysis for metric boundary data}
\label{sec:asymp}
In this section we will study the classical limit of the f\/irst order phase space path
integral~\eqref{eq:NCpwmetricreps} for the Ponzano--Regge model obtained through the non-commutative Fourier transform.
In particular, we will study the classical limit via the stationary phase approximation, f\/irst by using the usual
`commutative' variational method.
However, we discover that the resulting classical geometricity constraints on the classical metric variables depend on
the initial choice of quantization map~-- a~rather problematic outcome.
Therefore, we are compelled to adopt the \emph{non-commutative} variational method for the stationary phase
approximation in order to obtain the correct classical equations of motion, as in the analogous case of quantum
mechanics of a~point particle on ${\rm SO}(3)$, considered previously in~\cite{OR}.
We will again see that the non-commutative method leads to the correct and unambiguous classical geometricity
constraints on the simplicial metric variables, and of\/fer some further justif\/ication for the use of the
non-commutative variational calculus.
Moreover, the analysis shows how subtle the notion of ``classical limit'' is for the Ponzano--Regge amplitudes, which
are in the end convolutions of non-commutative planes waves, in the f\/lux representation.
We would expect similar subtleties to be relevant for 4d gravity models as well.

Let us begin by bringing the path integral~\eqref{eq:NCpwmetricreps} into a~form suitable for stationary phase
approximation via variational calculus.
We may use the expression~\eqref{eq:Eform} for the non-commutative plane wave to express~\eqref{eq:NCpwmetricreps} as
\begin{gather*}
\tilde{\mathcal{A}}_{\rm PR}(X_{ij}) = \int \bigg[\prod\limits_{(i,j)\in\mathcal{N}} \frac{\dd g_{ij}}{\kappa^3}\bigg]
\bigg[\prod\limits_{\substack{(i,j)\in\mathcal{N}
\\
i<j}}\frac{\dd Y_{ij}}{(2\pi\hbar\kappa)^3}\bigg] \bigg[\prod\limits_l\dd h_l\bigg]
\bigg[\prod\limits_{e\notin\partial\Delta}\frac{\dd Y_{e}}{(2\pi\hbar\kappa)^3}\bigg]
\\
\hphantom{\tilde{\mathcal{A}}_{\rm PR}(X_{ij}) =}{}
\times \bigg[\prod\limits_{e \notin \partial\Delta} c(H_{e^*}(h_l)) e^{\frac{i}{\hbar} Y_e \cdot
\zeta(H_{e^*}(h_l))}\bigg] \bigg[\prod\limits_{(i,j)\in\mathcal{N}} c\big(g_{ij}^{-1}\big) e^{\frac{i}{\hbar}X_{ij}\cdot
\zeta(g_{ij}^{-1})}\bigg]
\\
\hphantom{\tilde{\mathcal{A}}_{\rm PR}(X_{ij}) =}{}
\times \bigg[\prod\limits_{\substack{(i,j)\in\mathcal{N}
\\
i<j}} c\big(g_{ij} h_{j}^{-1} K_{ji}(h_l) h_{i} g_{ji}^{-1}\big) e^{\frac{i}{\hbar}Y_{ij}\cdot \zeta(g_{ij}
h_{j}^{-1}K_{ji}(h_l) h_{i} g_{ji}^{-1})} \bigg],
\end{gather*}
and by further combining the exponentials we obtain
\begin{gather}
\tilde{\mathcal{A}}_{\rm PR}(X_{ij}) = \int \bigg[\prod\limits_{(i,j)\in\mathcal{N}} \frac{\dd g_{ij}}{\kappa^3}
c\big(g_{ij}^{-1}\big)\bigg] \bigg[\prod\limits_{\substack{(i,j)\in\mathcal{N}
\\
i<j}}\frac{\dd Y_{ij}}{(2\pi\hbar\kappa)^3}\bigg] \bigg[\prod\limits_l\dd h_l\bigg]
\bigg[\prod\limits_{e\notin\partial\Delta}\frac{\dd Y_{e}}{(2\pi\hbar\kappa)^3}\bigg]
\nonumber
\\
\qquad
\times \bigg[\prod\limits_{e \notin \partial\Delta} c(H_{e^*}(h_l)) \bigg]
\bigg[\prod\limits_{\substack{(i,j)\in\mathcal{N}
\\
i<j}} c\big(g_{ij} h_{j}^{-1} K_{ji}(h_l) h_{i} g_{ji}^{-1}\big) \bigg]
\label{eq:PRamplitude}
\\
\times \exp\bigg\{\frac{i}{\hbar} \bigg[\sum\limits_{e \notin \partial\Delta} \! Y_e \cdot \zeta(H_{e^*}(h_l)) +
\!\sum\limits_{\substack{(i,j)\in\mathcal{N}
\\
i<j}} \! Y_{ij}\cdot \zeta\big(g_{ij} h_{j}^{-1} K_{ji}(h_l) h_{i} g_{ji}^{-1}\big) + \!\sum\limits_{(i,j)\in\mathcal{N}}\!  X_{ij}\cdot
\zeta(g_{ij}^{-1}) \bigg]\bigg\}.
\nonumber
\end{gather}
In this form the amplitude is amenable to a~stationary phase analysis through the study of the extrema of the
exponential
\begin{gather}
\mathcal{S}_{\rm PR}:= \sum\limits_{e \notin \partial\Delta} \! Y_e \cdot \zeta(H_{e^*}(h_l)) +
\sum\limits_{\substack{(i,j)\in\mathcal{N}
\\
i<j}} \! Y_{ij}\cdot \zeta\big(g_{ij} h_{j}^{-1} K_{ji}(h_l) h_{i} g_{ji}^{-1}\big)+\sum\limits_{(i,j)\in\mathcal{N}}\!
X_{ij}\cdot \zeta\big(g_{ij}^{-1}\big).\!\!\!
\label{eq:PRaction}
\end{gather}
We stress that this is just the classical action of discretized BF theory in its f\/irst order variables, the edge
vectors $Y_{e}$ and the parallel transports $h_l$, augmented by boundary terms.
Therefore, we expect to obtain in the classical limit the classical BF `equations of motion', that is, geometricity
constraints imposing f\/latness of holonomies around dual faces and closure of edge vectors for all triangles (up to the
appropriate parallel transports).

\subsection{Stationary phase approximation via commutative variational method}
We f\/irst proceed to consider the usual `commutative' stationary phase approximation of the f\/irst order
Ponzano--Regge path integral~\eqref{eq:NCpwmetricreps} by studying the extrema of the action~\eqref{eq:PRaction}.
There are f\/ive dif\/ferent kinds of integration variables in~\eqref{eq:NCpwmetricreps}: $Y_e$ for $e \notin
\partial\Delta$, $Y_{ij}$, $h_l$ in the bulk, $h_i$~touching the boundary and~$g_{ij}$, whose variations we will
consider in the following.
\begin{description}\itemsep=0pt
\item[Variation of $Y_e$:] Requiring the variation of the action with respect to $Y_e$ to vanish simply gives
\begin{gather*}
\zeta(H_{e^*}(h_l)) = 0 \ \Leftrightarrow \ H_{e^*}(h_l) = \mathbbm{1}
\end{gather*}
for all $e\notin\partial\Delta$, i.e., the f\/latness of the connection around the dual faces $e^*$ in the bulk.
\item[Variation of $Y_{ij}$:] Similarly, variation with respect to $Y_{ij}$ gives
\begin{gather*}
\zeta(g_{ij} h_{j}^{-1} K_{ji}(h_l) h_{i} g_{ji}^{-1}) = 0 \ \Leftrightarrow \ g_{ij} h_{j}^{-1} K_{ji}(h_l) h_{i}
g_{ji}^{-1} = \mathbbm{1}
\end{gather*}
for all $(i,j)\in\mathcal{N}$, $i<j$, i.e., the triviality of the connection around the faces $e^*$ dual to boundary
edges $e\in\partial\Delta$.
\item[Variation of $h_l$ in the bulk:] The variations for the group elements are slightly less trivial.
Taking left-invariant Lie derivatives of the exponential with respect to a~group element $h_{l'} \equiv g_{tf}$ in the
bulk, we obtain
\begin{gather*}
\sum\limits_{e \notin \partial\Delta} Y_e \cdot \mathcal{L}^{h_{l'}}_k\zeta(H_{e^*}(h_l)) +
\sum\limits_{\substack{(i,j)\in\mathcal{N}
\\
i<j}} Y_{ji}\cdot \mathcal{L}^{h_{l'}}_k\zeta\big(g_{ij} h_{j}^{-1} K_{ji}(h_l) h_{i} g_{ji}^{-1}\big) = 0
\qquad
\forall\, k.
\end{gather*}
Here, only the three terms in the sums depending on the holonomies around the boundaries of the three dual faces, which
contain $l':= tf$ are non-zero.
(Each dual edge $f^*$ belongs to exactly three dual faces $e^*$ of $\Delta^*$, since $\Delta^*$ is dual to
a~3-dimensional triangulation.) Now, using the fact uncovered through the previous variations that the holonomies around
the dual faces are trivial for the stationary phase conf\/igurations, and the property $\zeta(\ad_gh) = \Ad_g\zeta(h)$
of the coordinates, we obtain
\begin{gather*}
\sum\limits_{\substack{e \in \Delta
\\
e^*\ni f^*}} \epsilon_{fe}(\Ad_{G_{fe}} Y_e) = 0,
\end{gather*}
where $\Ad_{G_{fe}}$ implements the parallel transport from the frame of $Y_e$ to the frame of~$f$, and
$\epsilon_{fe}=\pm 1$ accounts for the orientation of $h_l$ with respect to the holonomy $H_{e^*}(h_l)$ and thus the
relative orientations of the edge vectors.
Clearly, this imposes the metric closure constraint for the three edge vectors of each bulk triangle $f \notin
\partial\Delta$ in the frame of~$f$.
This same condition gives the metric compatibility of the discrete connection, which in turn, if substituted back in the
classical action, before considering the other saddle point equations, turn the discrete 1st order action into the 2nd
order action for the triangulation~$\Delta$.
\item[Variation of $h_i$:] Varying a~$h_i$ we get
\begin{gather*}
\sum\limits_{\substack{(i,j)\in\mathcal{N}
\\
i<j}} Y_{ij}\cdot \mathcal{L}^{h_i}_k \zeta\big(g_{ij} h_{j}^{-1} K_{ji}(h_l) h_{i} g_{ji}^{-1}\big)\\
\qquad{}  +
\sum\limits_{\substack{(j,i)\in\mathcal{N}
\\
j<i}} Y_{ji}\cdot \mathcal{L}^{h_i}_k \zeta\big(g_{ji} h_{i}^{-1} K_{ij}(h_l) h_{j} g_{ij}^{-1}\big) = 0
\qquad
\forall\,  k.
\end{gather*}
Again there are three non-zero terms in this expression, which correspond to the boundary triangles
$f_j\in\partial\Delta$ neighboring $f_i$, i.e., such that $(i,j)\in\mathcal{N}$.
We obtain the closure of the boundary integration variables $Y_{ij}$ as
\begin{gather}
\sum\limits_{\substack{f_j\in\partial\Delta
\\
(i,j)\in\mathcal{N}}} \epsilon_{ji} \big(\Ad_{g_{ji}}^{-1}Y_{ij}\big) = 0,
\label{eq:Yboundclosure}
\end{gather}
where $\Ad_{g_{ji}}^{-1}$ parallel transports the edge vectors $Y_{ji}$ to the frame of the boundary triang\-le~$f_i$, and
$\epsilon_{ji}= \pm 1$ again accounts for the relative orientation.
\item[Variation of $g_{ij}$:] Taking Lie derivatives of the exponential with respect to a~$g_{ij}$, we obtain
\begin{gather}
Y_{ij}\cdot \mathcal{L}_k^{g_{ij}}\zeta\big(g_{ij} h_{j}^{-1} K_{ji}(h_l) h_{i} g_{ji}^{-1}\big) + X_{ij}\cdot
\mathcal{L}_k^{g_{ij}}\zeta\big(g_{ij}^{-1}\big) = 0
\qquad
\forall \, k
\nonumber
\\
\Leftrightarrow
\quad
\Ad_{g_{ij}}^{-1} Y_{ij} - D^{\zeta}(g_{ij}) X_{ij} = 0 = \Ad_{g_{ij}}^{-1} Y_{ij} + D^{\zeta}(g_{ji}) X_{ji}
\label{eq:Y=X}
\end{gather}
for all $i<j$, where we denote $(D^{\zeta}(g))_{kl}:= \tilde{\mathcal{L}}_k\zeta_l(g)$.
We see that this equation identif\/ies the boundary metric variables $X_{ij}$ with the integration variables $Y_{ij}$,
taking into account the orientation and the parallel transport between the frames of each vector, plus \emph{a
non-geometric deformation} given by the matrix $D^{\zeta}(g_{ij})$.\footnote{Also, in varying $g_{ij}$ we must assume
that the measure $c(g)\dd g$ on the group is continuous, which should be true for any reasonable choice of
a~quantization map, as it indeed is for all the cases we consider below.}
\end{description}

Thus, we have obtained the constraint equations corresponding to variations of all the integration variables.
In particular, by combining the equations~\eqref{eq:Y=X} with the boundary closure constraint~\eqref{eq:Yboundclosure},
we obtain
\begin{gather}
\label{eq:defXclosure}
\sum\limits_{\substack{f_j\in\partial\Delta
\\
(i,j)\in\mathcal{N}}} D^{\zeta}(g_{ij}) X_{ij} = 0
\qquad
\forall\, i,
\end{gather}
which gives, in general, a~\emph{deformed} closure constraint for the boundary metric edge variables~$X_{ij}$.
In addition, from~\eqref{eq:Y=X} alone we obtain a~\emph{deformed} identif\/ication
\begin{gather*}
\Ad_{g_{ij}} \big(D^\zeta(g_{ij})X_{ij}\big) = -\Ad_{g_{ji}} \big(D^\zeta(g_{ji})X_{ji}\big),
\end{gather*}
naturally up to a~parallel transport, of the boundary edge variables $X_{ij}$ and $X_{ji}$.
Accordingly, we obtain for the amplitude
\begin{gather}
\tilde{\mathcal{A}}_{\rm PR}(X_{ij}) \propto \int \bigg[\prod\limits_{(i,j)\in\mathcal{N}} \frac{\dd g_{ij}}{\kappa^3}
c\big(g_{ij}^{-1}\big)\bigg] \bigg[\prod\limits_{v\in\partial\Delta} \delta(H_v(g_{ij})) \bigg]
\bigg[\prod\limits_{f_i \in\partial\Delta}
\delta_\star\bigg(\sum\limits_{\substack{f_j\in\partial\Delta\\(i,j)\in\mathcal{N}}} D^{\zeta}(g_{ij}) X_{ij}\bigg) \bigg]
\nonumber
\\
\hphantom{\tilde{\mathcal{A}}_{\rm PR}(X_{ij}) \propto}{}
\star \bigg[\prod\limits_{\substack{(i,j)\in\mathcal{N}\\i<j}}
\delta_\star\bigg(\Ad_{g_{ij}} \big(D^\zeta(g_{ij})X_{ij}\big) + \Ad_{g_{ji}} \big(D^\zeta(g_{ji})X_{ji}\big)\bigg) \bigg]
\nonumber
\\
\hphantom{\tilde{\mathcal{A}}_{\rm PR}(X_{ij}) \propto}{}
\star
\exp\bigg\{\frac{i}{\hbar}\sum\limits_{(i,j)\in\mathcal{N}}X_{ij}\cdot\zeta\big(g_{ij}^{-1}\big)\bigg\}
\big(1+\mathcal{O}(\hbar)\big).
\label{eq:0thPRamp}
\end{gather}
The proportionality constant is given by the conf\/iguration space volume for the geometric conf\/igurations in the
bulk, which is generically inf\/inite but is cancelled by normalization.
The delta functions impose the constraints on boundary data discussed above.
In particular, $H_v(g_{ij})$~are the holonomies around the boundary vertices $v\in\partial\Delta$, whose triviality
follows from the triviality of the bulk holonomies.
Notice that one must write the integrand in terms of $\star$-products and $\star$-delta functions in order for the
constraints to be correctly imposed, since the amplitude acts on wave functions through $\star$-multiplication.
The exact form of the deformation matrix $D^\zeta_{kl}(g) \equiv \{X_k,\zeta_l\}(g) = \delta_{kl} +
\mathcal{O}(\kappa,|\ln(g)|)$, and accordingly the geometric content of these constraints, depends on the
coordinates~$\zeta$, which are determined by the discretization of the continuum BF action or, equivalently, the initial
choice of the quantization map.
We see that only in the abelian limit $\kappa\rightarrow 0$, $|\zeta|=\const$, do the dif\/ferent choices agree in
general, producing the undeformed discrete geometric constraints
\begin{gather*}
\sum\limits_{\substack{f_j\in\partial\Delta
\\
(i,j)\in\mathcal{N}}} X_{ij} = 0
\qquad
\forall \, f_i\in\partial\Delta
\qquad
\text{and}
\qquad
\Ad_{g_{ij}} X_{ij} = \Ad_{g_{ji}} X_{ji}
\qquad
\forall \, (i,j)\in\mathcal{N}
\end{gather*}
for the discretized boundary metric variables $X_{ij}\in\mathfrak{su}(2)^*$.

\subsubsection*{Some examples}
\label{subsec:examples}

Before we go on to consider the stationary phase boundary conf\/igurations obtained through the ordinary commutative
variational calculus for some specif\/ic choices of the coordinates~$\zeta$, and thus the associated quantization maps,
let us make a~few general remarks on the apparent dependence of the limit on this choice.
As we have already emphasized above, the exact functional form of the non-commutative plane waves, and thus the
coordinate choice, is determined ultimately by the choice of the quantization map and the $\star$-product that we thus
obtain.
We have found the general expression for the plane wave as a~$\star$-exponential
\begin{gather*}
E(g,X) = e_\star^{\frac{i}{\hbar\kappa}k(g)\cdot X} = \sum\limits_{n=0}^{\infty} \frac{1}{n!}
\left(\frac{i}{\hbar\kappa}\right)^n k^{i_1}(g) \cdots k^{i_n}(g) X_{i_1} \star \dots \star X_{i_n}.
\end{gather*}
From this expression we may observe that the way the Planck constant $\hbar$ enters into the plane wave is very subtle.
There are negative powers of $(\hbar\kappa)$ coming from the prefactor in the exponential, while from the
$\star$-monomials arise positive powers of $(\hbar\kappa)$.
The way these dif\/ferent contributions go together determines the explicit functional form of the non-commutative plane
wave, and accordingly the behavior in the classical limit $\hbar \rightarrow 0$.
Therefore, it is not too surprising that we may eventually f\/ind dif\/ferent classical limits for dif\/ferent choices
of plane waves through the application of the ordinary stationary phase method.
In particular, it is important to realize that the non-commutative plane wave itself is purely a~quantum object with an
ill-def\/ined classical limit, and therefore has no duty to coincide with anything in this limit.
For this reason, the stationary phase solutions corresponding to dif\/ferent $\star$-products may also dif\/fer from
each other, even though the $\star$-product itself coincides with the pointwise product in this limit.
On the contrary, in the abelian limit~$\kappa\rightarrow 0$ we also scale the coordinates~$k^i$ on the group, so that~$k^i/\kappa$ stay constant, since~$\kappa$ determines the scale associated to the group manifold.
Therefore, the non-commutative plane wave agrees with the usual commutative plane wave in this limit.
Only in the abelian limit may one expect the dif\/ferent choices of non-commutative structures lead to unambiguous
results, when one applies the commutative variational calculus.

{\it Symmetric {\rm \&} Duflo quantization maps.}
The symmetric quantization map $\mathfrak{Q}_{\tinyS}: \Pol(\mathfrak{su}(2)^*) \rightarrow U(\mathfrak{su}(2))$ is
determined by the symmetric operator ordering for monomials
\begin{gather*}
\mathfrak{Q}_{\tinyS}(X_{i_1}X_{i_2}\cdots X_{i_n}) \stackrel{!}{=} \frac{1}{n!}\sum\limits_{\sigma\in \Sigma_n}
\hat{X}_{i_{\sigma(1)}} \hat{X}_{i_{\sigma(2)}} \cdots \hat{X}_{i_{\sigma(n)}},
\end{gather*}
where $\Sigma_n$ is the group of permutations of~$n$ elements, and extends by linearity to any completion of the
polynomial algebra $\Pol(\mathfrak{su}(2)^*)$.
In particular, we have
\begin{gather*}
\mathfrak{Q}_{\tinyS}^{-1}(e^{\frac{i}{\hbar\kappa}k(g)\cdot \hat{X}}) = \sum\limits_{n=0}^{\infty}
\frac{i^n}{n!(\hbar\kappa)^n} k^{i_1}(g) \cdots k^{i_n}(g) \mathfrak{Q}_{\tinyS}^{-1}(\hat{X}_{i_1} \cdots \hat{X}_{i_n})
= e^{\frac{i}{\hbar\kappa}k(g)\cdot X} \equiv E_{\tinyS}(g,X)
\end{gather*}
and accordingly to this quantization prescription is associated a~non-commutative plane wave
with $c_{\tinyS}(g)=1$,
$\zeta_{\tinyS}(g) = -\frac{i}{\kappa}\ln(g) \in \mathfrak{su}(2)$, where the value of the logarithm is taken
in the principal branch~\cite{GOR}.

The Duf\/lo quantization map $\mathfrak{Q}_{\tinyD}$ is def\/ined as
\begin{gather*}
\mathfrak{Q}_{\tinyD} = \mathfrak{Q}_{\tinyS} \circ j^{\frac{1}{2}}(\vec{\partial}_x),
\end{gather*}
where $j^{\frac{1}{2}}(\vec{\partial}_x)$ is a~dif\/ferential operator associated to the function $j:\mathfrak{su}(2)
\rightarrow \mathbb{C}$ given~by
\begin{gather*}
j^{\frac{1}{2}}(X):= \det\left(\frac{\sinh(\frac{1}{2}\ad_X)}{\frac{1}{2}\ad_X} \right)^{\frac{1}{2}}.
\end{gather*}
The def\/inition is such that $\mathfrak{Q}_{\tinyD}$ restricts to an isomorphism from the~$g$-invariant functions on
$\mathfrak{su}(2)^*$ to~$g$-invariant operators (i.e., Casimirs) in $\overline{U(\mathfrak{su}(2))}$, and therefore the
Duf\/lo map can be considered as algebraically the most natural choice for a~quantization map.
In the Duf\/lo case we obtain $c_{\tinyD}(g)=\kappa|\zeta_{\tinyS}(g)|/\sin(\kappa|\zeta_{\tinyS}(g)|)$, but
the coordinates are the same $\zeta_{\tinyD} \equiv \zeta_{\tinyS}$ as for the symmetric quantization map, so the
amplitudes have the same stationary phase behavior in both cases.
In this respect it is important to note that even though the Duf\/lo factor
$c_{\tinyD}(g)=\kappa|\zeta_{\tinyS}(g)|/\sin(\kappa|\zeta_{\tinyS}(g)|)$
diverges for $\kappa|\zeta_{\tinyS}(g)| = \pi$, the
path integral measure is still well-behaved,
since $c_{\tinyD}(g)\dd g=(\sin(\kappa|\zeta_{\tinyS}|)/\kappa|\zeta_{\tinyS}|) \dd^3\zeta_{\tinyS}$ remains f\/inite.

We obtain from the equation~\eqref{eq:BCHder} in Appendix~\ref{sec:infBCH} for the deformation matrix as a~function of
the coordinates
\begin{gather}
\label{eq:DS}
D^{\tinyS}_{kl}(\zeta_{\tinyS}) = \left(\frac{\kappa|\zeta_{\tinyS}|}{\sin(\kappa|\zeta_{\tinyS}|)}\right)\!
\left[\cos(\kappa|\zeta_{\tinyS}|) \delta_{kl} + \left(\frac{\sin(\kappa|\zeta_{\tinyS}|)}{\kappa|\zeta_{\tinyS}|} -
\cos(\kappa|\zeta_{\tinyS}|) \right) \!\frac{\zeta_{\tinyS,k} \zeta_{\tinyS,l}}{|\zeta_{\tinyS}|^2} -
\kappa\epsilon_{klm}\zeta_{\tinyS}^m \right].\!\!\!\!
\end{gather}
This deformation matrix has the following nice property: $D_{kl}^{\tinyS}(k)k^l = k_k$.
This implies, in particular, that when the edge vectors are stable under the dual connection variables,
\mbox{$\Ad_{g_{ij}}X_{ij} = X_{ij}$} $\Leftrightarrow$ $k(g_{ij}) \propto X_{ij}$, we have $D^{\tinyS}(g_{ij})X_{ij} =
X_{ij}$, and therefore recover the undeformed closure constraints from~\eqref{eq:defXclosure}.
Accordingly, classical geometric boundary data with $\Ad_{g_{ij}}X_{ij} = X_{ij}$, $X_{ij} = -X_{ji}$ and
$\sum\limits_{j}X_{ij}=0$ satisf\/ies the constraint equations for the symmetric quantization map.
Except for the stability ansatz $\Ad_{g_{ij}}X_{ij} = X_{ij}$, however, there are undoubtedly other solutions to the
constraint equations that do not satisfy this stability requirement, but we have not explored the possibilities in this
general case.
It is nevertheless clear that these additional solutions do not correspond to simplicial geometries, since for them the
closure constraint is again deformed.

{\it Freidel--Livine--Majid quantization map.}
We will then consider the popular choice of Freidel--Livine--Majid quantization map~\cite{BDOT,BGO,BO,BO2,BO3, FL,FM,JMN},
which can be expressed in terms of the symmetric quantization map $\mathfrak{Q}_{\tinyS}$ and a~change of
parametrization $\chi: \mathfrak{su}(2) \rightarrow \mathfrak{su}(2)$ on ${\rm SU}(2)$ as~\cite{GOR}
\begin{gather*}
\mathfrak{Q}_{\tinyFLM} = \mathfrak{Q}_{\tinyS} \circ \chi,
\end{gather*}
where $\chi(k) = \frac{\sin^{-1}|k|}{|k|} k$.
The inverse coordinate transformation $\chi^{-1}(k) = \frac{\sin|k|}{|k|}k$, however, is two-to-one: it identif\/ies the
coordinates of the antipodes~$g$ and $-g$ as
\begin{gather*}
\chi^{-1}(k(g)) = \chi^{-1}\left(k(g) - \frac{\pi}{2}\frac{k(g)}{|k(g)|}\right) = \chi^{-1}(k(-g))
\qquad
\forall\, g\in {\rm SU}(2)\backslash \{e\}.
\end{gather*}
Therefore, the coordinates $\chi^{-1}(k)$ only cover the upper hemisphere ${\rm SU}(2)/\mathbb{Z}_2 \cong {\rm SO}(3)$, and the
resulting non-commutative Fourier transform is applicable only to functions on ${\rm SO}(3)$.

The FLM quantization map yields
\begin{gather*}
\mathcal{Q}_{\tinyFLM}^{-1}\big(e^{\frac{i}{\hbar\kappa} k(g)\cdot\hat{X}}\big)=e^{\frac{i}{\hbar\kappa}\frac{\sin|k(g)|}{|k(g)|}
k(g)\cdot X} \equiv E_{\tinyFLM}(g,X).
\end{gather*}
Accordingly, it leads to a~form of the non-commutative plane wave with $c_{\tinyFLM}(g)=1$,
$\zeta_{\tinyFLM}(g) = \frac{1}{\kappa}\frac{\sin|k(g)|}{|k(g)|} k(g) =
-\frac{i}{2\kappa}\tr_{\frac{1}{2}}(g\sigma^k)\sigma_k \in \mathfrak{su}(2)$, where $\tr_{\frac{1}{2}}$ denotes the
trace in the fundamental spin-$\frac{1}{2}$ representation of ${\rm SU}(2)$.
Due to the linearity of the trace, it is straightforward to calculate the deformation matrix
\begin{gather*}
D^{\tinyFLM}_{kl}(g)
= \frac{1}{2}\tr_{\frac{1}{2}}(g)\delta_{kl}+\frac{i}{2}\tr_{\frac{1}{2}}\big(g\sigma^j\big)\epsilon_{jkl}
\equiv\sqrt{1-\kappa^2 |\zeta_{\tinyFLM}(g)|^2}\delta_{kl}-\kappa\zeta_{\tinyFLM}^j(g) \epsilon_{jkl}.
\end{gather*}
Thus, according to our general description above, the classical discrete geometricity constraints are satisf\/ied by the
deformed boundary metric variables
\begin{gather*}
D^{\tinyFLM}(g_{ij})X_{ij} = \sqrt{1 - \kappa^2 |\zeta_{\tinyFLM}(g_{ij})|^2} X_{ij} - \kappa
(\zeta_{\tinyFLM}(g_{ij}) \wedge X_{ij}).
\end{gather*}
We have not solved these constraints explicitly, which would generically impose relations between the stationary phase
boundary connection $g_{ij}$ and the given boundary metric data $X_{ij}$.
However, one can easily conf\/irm that data corresponding to generic classical discrete geometries does not satisfy the
constraints, and therefore the geometry resulting from the constraints does not, in general, describe discrete
geometries.
In fact, the deformed and the undeformed closure constraints are compatible only for $g_{ij}\equiv \mathbbm{1}$, or
equivalently, in the abelian limit.
Therefore, we conclude that the non-commutative metric boundary variables do not have a~classical geometric
interpretation in the case of FLM quantization map outside the abelian approximation, when one studies the commutative
variation of the action.

\subsection{Stationary phase approximation via non-commutative\\ variational method}

We emphasize that in the above variation of the amplitude we did not take into account the deformation of phase space
structure, which appeared crucial for obtaining the correct classical equations of motion in~\cite{OR} in the case of
quantum mechanics of a~point particle on ${\rm SO}(3)$.
This could be guessed to be the origin of the discrepancies between the amplitudes corresponding to dif\/ferent choices
of quantization maps in the semi-classical limit.
Indeed, we may def\/ine the \emph{non-commutative variation} $\delta_\star S$ of the action~$S$ in the amplitude via
\begin{gather*}
e_\star^{i\delta_\star S + \mathcal{O}(\delta^2)} \equiv e_\star^{-iS} \star e_\star^{iS^\delta},
\end{gather*}
where the $\star$-product acts on the f\/ixed boundary variables $X_{ij}$, $\mathcal{O}(\delta^2)$ refers to terms
higher than f\/irst order in the variations, and $S^\delta\equiv S(g_{ij}\delta g_{ij},X_{ij} + \delta X_{ij})$ is the
varied action.
It is easy to see that the non-commutative variation so def\/ined undeforms the above identif\/ication~\eqref{eq:Y=X} of~$X_{ij}$ and~$Y_{ij}$ (up to orientation and parallel transport), simply because we have
\begin{gather*}
E\big(g^{-1},X\big) \star E\big(ge^{i\epsilon Z},X\big) = \underbrace{E\big(g^{-1},X\big) \star E(g,X)}_{=1} \star E\big(e^{i\epsilon Z},X\big) =
e^{\frac{i}{\hbar\kappa}\epsilon (Z\cdot X) + \mathcal{O} (\epsilon^2 )}
\end{gather*}
for any $Z\in \mathfrak{su}(2)$ and $\epsilon \in \mathbb{R}$ implementing the variation of~$g$, so that $\delta g =
e^{i\epsilon Z}$.
Then, all the above results for variations remain the same by requiring the non-commutative variation $\delta_\star S$
of the action to vanish except for equation~\eqref{eq:Y=X}, which becomes undeformed, i.e., we obtain the geometric
identif\/ication $\Ad_{g_{ij}}X_{ij} = Y_{ij} = -\Ad_{g_{ji}}X_{ji}$.
Thus, the non-geometric deformation of the constraints does not appear, and we recover exactly the simplicial geometry
relations for the boundary metric variables, regardless of the choice of a~quantization map.

The non-commutative geometric interpretation of the leading order contribution to the amplitude obtained through the
non-commutative variation is largely an open question at the moment~-- and a~very interesting one as well.
Clearly, the non-commutative leading order is dif\/ferent from the ordinary commutative result, because we are not
considering the usual commutative limit, but another kind of limit that takes into account the non-commutative structure
of the phase space.
Indeed, this dif\/ference is more than welcome, because the commutative result depends on the choice of a~quantization
map, which is unacceptable, as we have emphasized.
Our calculations below show, in fact, that the application of the commutative variational method to an integral kernel
that is a~function of non-commutative variables leads to a~result that does not represent the leading order in $\hbar$:
We conf\/irm in Section~\ref{sec:6j} that the results obtained (only) by the non-commutative stationary phase analysis
agree with those obtained through the indisputable commutative analysis in the coherent state representation.
We still lack a~complete understanding of the non-commutative variations, but we suspect that the need for the
non-commutative variational method arises, because the amplitude $\tilde{\mathcal{A}}_{\rm PR}(X_{ij})$ acts as the integral
kernel for the propagator with respect to the $\star$-product and not the pointwise product.
As the $\star$-product itself exhibits $\hbar$-dependence, this may modify the asymptotic behavior.
The classical constraint equations that we recover via the non-commutative variations are presumably the ones that are
imposed on the boundary states by the propagator (again, acting with the $\star$-product) in the classical limit.
However, this needs to be substantiated by further research.

To begin with, let us consider the partially `of\/f-shell' amplitude, where we only substitute the identif\/ications
$\Ad_{g_{ij}}X_{ij} = Y_{ij} = -\Ad_{g_{ji}}X_{ji}$ arising from the variations of the boundary connection $g_{ij}$.
The substitution is done, again, by multiplying the amplitude by $\star$-delta functions imposing the identities, and
integrating over $Y_{ij}$.
We also integrate over $X_{ji}$ for $i<j$ in order to explicitly impose the identif\/ications $\Ad_{g_{ij}}X_{ij} =
-\Ad_{g_{ji}}X_{ji}$ on the boundary variables.
Using the properties of the non-commutative plane waves, and denoting by $\tilde{\mathcal{A}}_{\rm PR}^{\star {\rm lo}}(X_{ij})$
the leading order contribution in $\hbar$ to the amplitude obtained via the non-commutative stationary phase method, we
f\/ind from~\eqref{eq:PRamplitude} through a~simple substitution
\begin{gather}
\tilde{\mathcal{A}}_{\rm PR}^{\star {\rm lo}}(X_{ij}) \propto \!\int\! \bigg[\prod\limits_{\substack{(i,j)\in\mathcal{N}
\\
i<j}}\! \frac{\dd g_{ij}}{\kappa^3} c\big(g_{ij}^{-1}\big) c\big(g_{ij}^{-1}g_{ji}g_{ij}\big) \bigg] \bigg[\prod\limits_l\dd h_l\bigg]
\bigg[\prod\limits_{e\notin\partial\Delta}\!\frac{\dd Y_{e}}{(2\pi\hbar\kappa)^3}\bigg] \bigg[\prod\limits_{e \notin
\partial\Delta}\! c(H_{e^*}(h_l)) \bigg]
\nonumber
\\
\hphantom{\tilde{\mathcal{A}}_{\rm PR}^{\star {\rm lo}}(X_{ij}) \propto}{}
\times \bigg[\prod\limits_{\substack{(i,j)\in\mathcal{N}
\\
i<j}} c\big(h_{j}^{-1} K_{ji}(h_l) h_{i} g_{ji}^{-1}g_{ij}\big) \bigg] \exp\bigg\{\frac{i}{\hbar} \sum\limits_{e \notin
\partial\Delta} Y_e \cdot \zeta(H_{e^*}(h_l))\bigg\}
\nonumber
\\
\hphantom{\tilde{\mathcal{A}}_{\rm PR}^{\star {\rm lo}}(X_{ij}) \propto}{}
\times \exp\bigg\{\sum\limits_{\substack{(i,j)\in\mathcal{N}
\\
i<j}} X_{ij}\cdot \zeta\big(h_{j}^{-1} K_{ji}(h_l) h_{i} g_{ji}^{-1}g_{ij}\big)\bigg\}
\nonumber
\\
\hphantom{\tilde{\mathcal{A}}_{\rm PR}^{\star {\rm lo}}(X_{ij}) \propto}{}
\star \exp\bigg\{\sum\limits_{\substack{(i,j)\in\mathcal{N}
\\
i<j}} X_{ij}\cdot \zeta\big(g_{ij}^{-1}g_{ji}g_{ij}\big)\bigg\} \star \exp\bigg\{\sum\limits_{\substack{(i,j)\in\mathcal{N}
\\
i<j}} X_{ij}\cdot \zeta\big(g_{ij}^{-1}\big)\bigg\}
\nonumber
\\
\hphantom{\tilde{\mathcal{A}}_{\rm PR}^{\star {\rm lo}}(X_{ij}) }{}
= \int \bigg[\prod\limits_l\dd h_l\bigg] \bigg[\prod\limits_{e\notin\partial\Delta}\dd Y_{e}\bigg]
\bigg[\prod\limits_{\substack{(i,j)\in\mathcal{N}
\\
i<j}} c\big(h_{j}^{-1} K_{ji}(h_l) h_{i}\big) \bigg] \bigg[\prod\limits_{e \notin \partial\Delta} c(H_{e^*}(h_l)) \bigg]
\nonumber
\\
\hphantom{\tilde{\mathcal{A}}_{\rm PR}^{\star {\rm lo}}(X_{ij}) \propto}{}
\times \exp\bigg\{\frac{i}{\hbar} \bigg[\sum\limits_{e \notin \partial\Delta} Y_e \cdot \zeta(H_{e^*}(h_l)) +
\sum\limits_{\substack{(i,j)\in\mathcal{N}
\\
i<j}} X_{ij}\cdot \zeta\big(h_{j}^{-1} K_{ji}(h_l) h_{i}\big) \bigg]\bigg\}.
\label{eq:whoop}
\end{gather}
In fact, there is a~subtlety in this calculation in choosing the correct ordering of the non-commutative plane waves,
which depend on the same non-commutative edge vector after integrating over the $\star$-delta functions, in the f\/irst
expression of~\eqref{eq:whoop}.
We were guided here in the choice by the appropriate geometric form of the result.
Indeed, the exponential clearly ref\/lects the typical structure of a~3d discrete gravity action: (i)~It contains bulk
terms $Y_e \cdot \zeta(H_{e^*}(h_l))$, which couple bulk edge vectors and the holonomies around the dual faces, thus
associated with def\/icit angles.
(ii)~It has boundary terms $X_{ij}\cdot \zeta(h_{j}^{-1} K_{ji}(h_l) h_{i})$, which couple boundary edge vectors with
the group elements that represent parallel transports between adjacent boundary triangles to the edge, thus associated
with dihedral angles.

To make the connection to Regge calculus even clearer, let us adopt the non-commutative structure arising from the
symmetric quantization map, and set $H_{e^*}(h_l) \equiv \exp(i\theta_e \hat{n}_e\cdot\vec{\sigma})$ and $h_{j}^{-1}
K_{ji}(h_l) h_{i} \equiv \exp(i\theta_{ji}\hat{n}_{ji}\cdot\vec{\sigma})$ in the spin-$\frac{1}{2}$ representation,
where $\theta_e,\theta_{ji}\in [0,\pi]$ are now the class angles of the group elements and $\hat{n}_e,\hat{n}_{ji}\in
S^2$ unit vectors.
Then, we may write
\begin{gather*}
Y_e \cdot \zeta_{\tinyS}(H_{e^*}(h_l)) = |Y_e| \frac{\theta_e}{\kappa} \left(\frac{Y_e}{|Y_e|}\cdot\hat{n}_e\right),
\qquad
X_{ij}\cdot \zeta\big(h_{j}^{-1} K_{ji}(h_l) h_{i}\big) =
|X_{ij}|\frac{\theta_{ji}}{\kappa}\left(\frac{X_{ij}}{|X_{ij}|}\cdot\hat{n}_{ji}\right).
\end{gather*}
Considering then variations in the unit vectors $\hat{n}_e$ and $\hat{n}_{ji}$, it is immediate to f\/ind that the
stationary phase of the amplitude is given by $\hat{n}_e = \pm \frac{Y_e}{|Y_e|}\cdot\hat{n}_e$ and $\hat{n}_{ji} = \pm
\frac{X_{ij}}{|X_{ij}|}$, the signs corresponding to the two opposite orientations of the edge vectors or, equivalently,
the dual faces.
Now, if a~conf\/iguration of edge vectors satisf\/ies the constraints for a~given discrete connection, it does so also
for the oppositely oriented conf\/iguration obtained by reversing the orientations of all the dual faces.
For the oppositely oriented conf\/iguration the holonomies around dual faces are also inverted, which gives opposite
signs for $\hat{n}_e$ and $\hat{n}_{ji}$ with respect to the original conf\/iguration.
Therefore, choosing $\hat{n}_e$ and $\hat{n}_{ji}$ to have positive signs for one of the orientations and thus negative
signs for the other, we may further write for the Ponzano--Regge amplitude in the semi-classical limit
\begin{gather*}
\tilde{\mathcal{A}}_{\rm PR}^{\star {\rm lo}}(X_{ij}) \propto \int \bigg[\prod\limits_l\dd h_l\bigg]
\bigg[\prod\limits_{e\notin\partial\Delta}\dd Y_{e}\bigg] \cos\bigg(\frac{i}{\hbar\kappa} \bigg[\sum\limits_{e \notin
\partial\Delta} |Y_e| \theta_e + \sum\limits_{\substack{(i,j)\in\mathcal{N}
\\
i<j}} |X_{ij}|\theta_{ji} \bigg]\bigg),
\end{gather*}
where the cosine arises from summing the contributions from both orientations of the triangulation.
The argument of the cosine is exactly the f\/irst order Regge action, well-known from discrete gravity.
Notice, however, that the def\/icit angles $\theta_e$ and the dihedral angles~$\theta_{ji}$ still depend on the discrete
bulk connection given by the group elements~$h_l$, which are integrated over in the amplitude.
Also, we have not yet imposed the closure constraints on the edge vectors, which arise from the variations of the bulk
connection.
These constraints impose the closure of edge vectors for each triangle, taking account orientations and parallel
transports.
At the same time they impose the metricity of the discrete connection and restrict the integrals over $Y_e$ to the
geometric conf\/igurations, as in Regge gravity\footnote{We note that for some choices of a~quantization map, such as
the Duf\/lo map, the $\star$-delta function does not depend on a~simple linear combination of its arguments, and thus
the closure constraints must be non-linear outside the strict classical regime to match the exact
result~\eqref{eq:metricBF}.
However, for the symmetric quantization map we consider here the closure constraints remain undeformed in all orders.}.
Solving for the discrete connection $h_l$ in terms of the edge vectors from the constraint equations (when possible,
i.e., for non-degenerate conf\/igurations) leads to the second order Regge action and to the form
\begin{gather*}
\tilde{\mathcal{A}}_{\rm PR}^{\star {\rm lo}}(X_{ij}) \propto \bigg[\prod\limits_{e\notin\partial\Delta}\dd Y_{e}\bigg]
\cos\bigg(\frac{i}{\hbar\kappa} \bigg[\sum\limits_{e \notin \partial\Delta} |Y_e| \theta_e +
\sum\limits_{\substack{(i,j)\in\mathcal{N}
\\
i<j}} |X_{ij}|\theta_{ji} \bigg]\bigg)
\end{gather*}
for the Ponzano--Regge amplitude, where now the def\/icit and dihedral angles are functions of the edge vectors, and
only geometric conf\/igurations of the edge vectors are integrated over.
Finally, we emphasize that for other choices of non-commutative structures, other than the one associated with the
symmetric quantization map, we may obtain more complicated dependence on the dihedral and def\/icit angles.
For example, the Freidel--Livine--Majid map leads to the compactif\/ied Regge action considered
in~\cite{Caselle89,Kawamoto91}.
However, all choices of non-commutative structures result in the same form as above in the regime of small def\/icit and
dihedral angles.
We thus see that the Regge action naturally arises in the semi-classical limit of the Ponzano--Regge model in terms of
the proper phase space variables.

Let us then move on to consider the `on-shell' case, where we impose all the classical constraints on the path integral
arising from the (non-commutative) stationary phase analysis.
In this case the leading order semi-classical contribution to the Ponzano--Regge amplitude~\eqref{eq:0thPRamp} reads in
detail before integrating out the bulk variables
\begin{gather*}
\tilde{\mathcal{A}}_{\rm PR}^{\star {\rm lo}}(X_{ij})
\propto
\int \bigg[\prod\limits_{\substack{(i,j)\in\mathcal{N}
\\
i<j}}\dd g_{ij}\dd g_{ji} \, c\big(g_{ij}^{-1}g_{ji}\big)\bigg]
\bigg[\prod\limits_l \dd h_l\bigg]
\bigg[\prod\limits_{e\notin\partial\Delta} \dd X_e\bigg] \Bigg[\prod\limits_{e\notin\partial\Delta}
\delta(H_{e^*}(h_l))\Bigg]
\nonumber
\\
\hphantom{\tilde{\mathcal{A}}_{\rm PR}^{\star {\rm lo}}(X_{ij})\propto}{}
\times \bigg[\prod\limits_{\substack{(i,j)\in\mathcal{N}
\\
i<j}} \delta\big(g_{ij}h_j^{-1}K_{ji}(h_l)h_ig_{ji}^{-1}\big)\bigg] \bigg[\prod\limits_{f\notin\partial\Delta}
\delta_\star\bigg(\sum\limits_{e\in f} \epsilon_{fe}\Ad_{h_{fe}}X_e\bigg) \bigg]
\nonumber
\\
\hphantom{\tilde{\mathcal{A}}_{\rm PR}^{\star {\rm lo}}(X_{ij})\propto}{}
\star \bigg[\prod\limits_i \delta_\star\bigg(\sum\limits_{\substack{(i,j)\in\mathcal{N}
\\
j>i}}\! X_{ij} - \sum\limits_{\substack{(i,j)\in\mathcal{N}
\\
j<i}} \!\Ad_{g_{ij}^{-1}g_{ji}}X_{ji}\bigg)\bigg] \star \exp\bigg\{\frac{i}{\hbar} \sum\limits_{\substack{(i,j)\in\mathcal{N}
\\
i<j}}\! X_{ij} \cdot \zeta\big(g_{ij}^{-1}g_{ji}\big) \!\bigg\},
\end{gather*}
where we have identif\/ied $Y_{ij}:= \Ad_{g_{ij}}X_{ij} = -\Ad_{g_{ji}}X_{ji}$ for all $(i,j)\in\mathcal{N}$ such that
$i<j$.
Here, the delta functions on the group impose triviality of the holonomies, which implies f\/latness of the discrete
connection.
The $\star$-delta functions impose closure of the edge vectors $e\in f$ belonging to the bulk and boundary triangles
$f\in\Delta$, which in the discrete gravity context corresponds to the metricity of the discrete connection.
The action is reduced due to the imposition of the f\/latness constraints to a~simple boundary term.

Since the amplitude depends only on $G_{ij}$ and not the individual $g_{ij}$, we may further apply a~change of variables
by denoting $G_{ij}:= g_{ji}^{-1}g_{ij}$ and $G_{ji} = G_{ij}^{-1}$ for all $i<j$.
These are the group elements that represent parallel transports between centers of boundary triangles, and are therefore
naturally related to the dihedral angles of Regge calculus.
By also integrating over the bulk variables, we obtain
\begin{gather}
\tilde{\mathcal{A}}_{\rm PR}^{\star {\rm lo}}(X_{ij})
\propto
\int \bigg[\prod\limits_{\substack{(i,j)\in\mathcal{N}
\\
i<j}}\dd G_{ij}\,  c(G_{ij})\bigg] \bigg[\prod\limits_{v\in\partial\Delta} \delta(H_v(G_{ij})) \bigg]
\nonumber\\
\hphantom{\tilde{\mathcal{A}}_{\rm PR}^{\star {\rm lo}}(X_{ij})\propto}{}
\times
\bigg[\prod\limits_i
\delta_\star\bigg(\sum\limits_{\substack{(i,j)\in\mathcal{N}
\\
j>i}} X_{ij} - \sum\limits_{\substack{(i,j)\in\mathcal{N}
\\
j<i}} \Ad_{G_{ij}}^{-1}X_{ji}\bigg)\bigg]
\nonumber
\\
\hphantom{\tilde{\mathcal{A}}_{\rm PR}^{\star {\rm lo}}(X_{ij})\propto}{}
\star \exp\bigg\{\frac{i}{\hbar} \sum\limits_{\substack{(i,j)\in\mathcal{N}
\\
i<j}} X_{ij} \cdot \zeta\big(G_{ij}^{-1}\big) \bigg\}.
\label{eq:PRG}
\end{gather}
The integrations over the bulk connection result in the delta functions imposing f\/latness of the boundary holonomies
$H_v(G_{ij})$ around all boundary vertices $v\in\partial\Delta$.
In addition, the integrations over the bulk variables yield the volume of geometric bulk conf\/igurations, which
contributes only to the normalization factor.
The result of the calculation is exactly as we would expect from a~f\/irst order 3d discrete gravity action, namely, it
is expressed as a~function of edge vectors, where parallel transports are integrated over f\/lat connections, which
allow the edge vectors to satisfy the closure constraints.

One may further decompose the integrals in~\eqref{eq:PRG} over group elements $G_{ij}$ into integrals over dihedral
class angles $\theta_{ij}:= |k(G_{ij})|$, where $k(g):=-i\ln(g) \in \mathfrak{su}(2) \cong \mathbb{R}^3$, and integrals
over unit vectors $\hat{n}_{ij}:= k(G_{ij})/|k(G_{ij})| \in S^2$, such that $G_{ij} \equiv \exp(i\theta_{ij}
\hat{n}_{ij}\cdot \vec{\sigma})$ in the spin-$\frac{1}{2}$ representation.
Adopting again the symmetric quantization map, we then have for the exponent
\begin{gather*}
\sum\limits_{\substack{(i,j)\in\mathcal{N}
\\
i<j}} X_{ij} \cdot \zeta_{\tinyS}(G_{ij}^{-1}) = - \frac{1}{\kappa}\sum\limits_{\substack{(i,j)\in\mathcal{N}
\\
i<j}} |X_{ij}| \theta_{ij} \left(\frac{X_{ij}}{|X_{ij}|}\right) \cdot \hat{n}_{ij}.
\end{gather*}
The stationary phase with respect to the integrals over $\hat{n}_{ij}\in S^2$ is given by $\frac{X_{ij}}{|X_{ij}|} \cdot
\hat{n}_{ij} = \pm 1$.
Now the two solutions correspond to opposite orientations of the boundary, both contributing to the dominant phase with
opposite signs, again turning the exponential into a~cosine.
Thus, we obtain
\begin{gather}
\tilde{\mathcal{A}}_{\rm PR}^{{\rm lo}}(X_{ij}) \propto \int \Bigg[\prod\limits_{\substack{(i,j)\in\mathcal{N}
\\
i<j}} \dd \theta_{ij}\left(\frac{\sin\theta_{ij}}{\theta_{ij}}\right)^2 \Bigg] \bigg[\prod\limits_{v\in\partial\Delta}
\delta(H_v(G_{ij})) \bigg]
\nonumber
\\
\hphantom{\tilde{\mathcal{A}}_{\rm PR}^{{\rm lo}}(X_{ij}) \propto}{}
\times \bigg[\prod\limits_i \delta_\star\bigg(\sum\limits_{\substack{(i,j)\in\mathcal{N}
\\
j>i}} X_{ij} - \sum\limits_{\substack{(i,j)\in\mathcal{N}
\\
j<i}} \Ad_{G_{ij}}^{-1}X_{ji}\bigg) \bigg] \star \cos\bigg(\frac{i}{\hbar\kappa} \sum\limits_{\substack{(i,j)\in\mathcal{N}
\\
i<j}} |X_{ij}| \theta_{ij}\bigg),\!\!\!
\label{eq:PRregge}
\end{gather}
where $G_{ij} \equiv \exp(-i\theta_{ij} (X_{ij}/|X_{ij}|)\cdot\vec{\sigma})$ and the $(\sin\theta_{ij}/\theta_{ij})^2$
factors arise from the Haar measure.
The second order expression may be obtained by solving the constraint equations for the boundary dihedral angles
$\theta_{ij}$ in terms of the edge vectors $X_{ij}$, and substituting into the above formula.
This leads to an expression of the form
\begin{gather*}
\tilde{\mathcal{A}}_{\rm PR}^{{\rm lo}}(X_{ij}) \propto \cos\bigg(\frac{i}{\hbar\kappa} \sum\limits_{\substack{i,j
\\
i<j}} |X_{ij}| \theta_{ij}(X_{kl})\bigg).
\end{gather*}
This is the second order Regge action in terms of only the boundary metric data.

Thus, we have verif\/ied that the non-commutative variational method for the stationary phase approximation leads to the
correct classical geometric constraints for the Ponzano--Regge model and agrees with the exact analysis of the
amplitude.
A~similar result was obtained by us for the dual non-commutative representation of quantum mechanics on the group
${\rm SO}(3)$ previously in~\cite{OR}.
The use of such a~non-commutative variational method may be motivated by noting that, in calculating transition
amplitudes for boundary states, $\tilde{\mathcal{A}}_{\rm PR}(X_{ij})$ acts as an integral kernel with respect to the
$\star$-product, and not with respect to the commutative pointwise product.
However, more work on this point is certainly needed in order to achieve a~better understanding that would def\/initely
settle the issue.
Granting the use of the non-commutative stationary phase method, we have thus shown that the phase space path integral
for the Ponzano--Regge model obtained through the non-commutative Fourier transform facilitates a~straigthforward
asymptotic analysis of the classical limit of the model.
We also showed that the semi-classical approximation to the amplitude is given by the cosine of the Regge action, in
agreement with previous studies~\cite{DGH}.

Finally, we would like to comment on our saddle point analysis in comparison with the existing ones in the
literature~\cite{CF, DGH}.
Besides not having a~real-valued contribution to the exponent (that is, an imaginary contribution to the classical
action), and thus having a~simpler variational principle given by a~standard (albeit non-commutative) saddle point
approximation, we also do not have any logarithmic term.
In other words, the f\/lux representation of the spin foam model gives a~discrete BF path integral in terms of the
standard classical action.
The choice of quantization map af\/fects the exact form of such discrete classical action, but in a~rather minor way.
In fact, it leads to a~discretization of the continuum curvature in terms of either the holonomy of the connection (a
group element), for the FLM map, or its logarithm (a Lie algebra element), for the Duf\/lo map.
In general, the analysis appears to be more straightforward, than the one based on the spin (or coherent state)
representation, as one would expect from a~straightforward path integral representation.
We will exemplify this comparison in the simple case of a~triangulation formed by a~single tetrahedron.

\section{Semi-classical limit for a~tetrahedron}
\label{sec:6j}

To conclude our asymptotic analysis, we consider in this section the relation of the classical limit of Ponzano--Regge
amplitudes in terms of non-commutative boundary metric variables to the usual formulation of spin foam asymptotics as
the large spin limit in the spin representation.
We will restrict our treatment to the case of a~single tetrahedron, since in this case the asymptotics for the
Ponzano--Regge amplitude is simple and well-known in the literature.
In particular, it has been found that (for non-degenerate boundary data) the amplitude of a~tetrahedron is approximated
in the large spin limit by the cosine of the Regge action~\cite{DGH,KaminskiSteinhaus13}.
We derive this result from the asymptotic behavior obtained through the non-commutative phase space path integral, and
thus establish a~f\/irm connection between the two asymptotic analyses.

For a~single tetrahedron the Ponzano--Regge amplitude with boundary connection data reads
\begin{gather}
\label{eq:PRtetra}
\mathcal{A}_{\rm PR}(g_{ij}) = \int \bigg[\prod\limits_i \dd h_i\bigg] \prod\limits_{\substack{i,j
\\
i<j}} \delta(g_{ij}h_j^{-1}h_ig_{ji}^{-1})
\end{gather}
where $i,j=1,\ldots,4$ label the boundary triangles, and the notation is chosen as in
Fig.~\ref{fig:treeholos}.\footnote{Of course, in the case of a~single tetrahedron $K_{ij}=\mathbbm{1}$ for all~$i$,~$j$,
since the paths along which $K_{ij}$ parallel transport (as in Fig.~\ref{fig:treeholos}) are of length~0.} The deltas
impose f\/latness of the connection around all six wedges of the tetrahedra
(see Fig.~\ref{fig:tetrasubdiv}, where wedges are illustrated in grey). Now, this amplitude may be transformed into
other representations described by other types of boundary data using dif\/ferent bases of functions on~${\rm SU}(2)$.
In particular, as already explained above, one may transform the amplitude into the basis given by the non-commutative
plane waves, in which case the amplitude is expressed as a~function of non-commutative boundary edge vectors.
Adopting for convenience the symmetric quantization map, this yields
\begin{gather*}
\tilde{\mathcal{A}}_{\rm PR}(X_{ij}) = \int \bigg[\prod\limits_{i,j} \dd g_{ij} E(g_{ij},X_{ij}) \bigg]
\mathcal{A}_{\rm PR}(g_{ij}) = \int \bigg[\prod\limits_{i,j} \dd g_{ij}\bigg] \bigg[\prod\limits_{i} \dd h_{i}\bigg]
\bigg[\prod\limits_{\substack{i,j
\\
i<j}} \dd Y_{ji}\bigg]
\nonumber
\\
\hphantom{\tilde{\mathcal{A}}_{\rm PR}(X_{ij}) =}{}
\times \exp\bigg\{\frac{i}{\hbar} \bigg(\sum\limits_{\substack{i,j
\\
i<j}} Y_{ij} \cdot \zeta_{\tinyS}\big(g_{ij}h_j^{-1}h_ig_{ji}^{-1}\big) - \sum\limits_{i,j} X_{ij} \cdot
\zeta_{\tinyS}(g_{ij})\bigg)\bigg\}.
\end{gather*}
We emphasize that this is again nothing but the f\/irst order action for discrete 3d gravity with boundary terms.
Let us brief\/ly sketch the asymptotic analysis for the Ponzano--Regge amplitude for a~tetrahedron in non-commutative
boundary metric variables.
As before, by studying the non-commutative variations of the action
\begin{gather*}
\mathcal{S}_{\rm PR} = \sum\limits_{\substack{i,j
\\
i<j}} Y_{ij} \cdot \zeta_{\tinyS}\big(g_{ij}h_j^{-1}h_ig_{ji}^{-1}\big) - \sum\limits_{i,j} X_{ij} \cdot \zeta_{\tinyS}(g_{ij})
\end{gather*}
we recover as stationary phase solutions in the classical limit the constraint equations
\begin{itemize}\itemsep=0pt
\item Wedge f\/latness\footnote{This is basically the condition enforcing the piece-wise f\/lat situation, i.e., the
f\/latness of the tetrahedron.}: $g_{ij}h_j^{-1}h_ig_{ji}^{-1} = \mathbbm{1}$ for all $i<j$,
\item Identif\/ication of boundary edge vectors: $\Ad_{g_{ij}} X_{ij} = Y_{ij} = -\Ad_{g_{ji}} X_{ji}$ for all $i<j$,
and
\item Closure of the boundary edge vectors: $\sum\limits_{j\neq i} \epsilon_{ij}X_{ji} = 0$ for all~$i$.
\end{itemize}
We then substitute these identities into the amplitude, whereby we f\/ind
\begin{gather}
\tilde{\mathcal{A}}_{\rm PR}^{{\rm lo}}(X_{ij}) \propto \int \bigg[\prod\limits_{i,j} \dd g_{ij}\bigg]
\bigg[\prod\limits_{v\in\partial\Delta} \delta(H_v(g_{ij})) \bigg] \bigg[\prod\limits_i \delta_\star\bigg(\sum\limits_{j> i}
X_{ij} - \sum\limits_{j<i} \Ad_{g_{ij}^{-1}g_{ji}} X_{ji}\bigg) \bigg]
\nonumber
\\
\hphantom{\tilde{\mathcal{A}}_{\rm PR}^{{\rm lo}}(X_{ij}) \propto}{}
\star \exp\bigg\{\frac{i}{\hbar} \sum\limits_{\substack{i,j
\\
i<j}} X_{ij} \cdot \zeta_{\tinyS}\big(g_{ij}^{-1}g_{ji}\big)\bigg\}
\label{eq:PRtetrag}
\end{gather}
as the leading order contribution in $\hbar$ to the amplitude.
Here, $\delta(H_v(g_{ij}))$ impose f\/latness of holonomies around all boundary vertices $v\in\partial\Delta$ of the
tetrahedron, which arise from the delta functions $\delta(g_{ij}h_j^{-1}h_ig_{ji}^{-1})$ by integrating over all $h_i$.
More precisely, we obtain the constraint only for three of the four vertices, which already imposes f\/latness for the
fourth vertex as well.
These conditions are nothing else than the Hamiltonian constraint of 3d gravity, generating the simplicial
dif\/feomorphisms at each vertex of the triangulation
(see, for example,~\cite{Bahr:2009ku,BGO}). We may further apply a~change of variables
by denoting $G_{ij}:=g_{ji}^{-1}g_{ij}$ and $G_{ji} = G_{ij}^{-1}$ for all $i<j$.
Again, these are the group elements that represent parallel transports between centers of boundary triangles, and are
therefore naturally related to the dihedral angles of Regge calculus.
As before in the more general case, one may further decompose the integrals over group elements $G_{ij}$ into integrals
over dihedral class angles $\theta_{ij}:= \kappa |\zeta_{\tinyS}(G_{ij})|$ and integrals over unit vectors
$\hat{n}_{ij}:= \zeta_{\tinyS}(G_{ij})/|\zeta_{\tinyS}(G_{ij})| \in S^2$.
The stationary phase equations for the unit vectors $\hat{n}_{ij}$ lead to the expression
\begin{gather}
\tilde{\mathcal{A}}_{\rm PR}^{{\rm lo}}(X_{ij}) \propto \!\int\! \bigg[\prod\limits_{\substack{i,j
\\
i<j}} \dd \theta_{ij}\left(\frac{\sin\theta_{ij}}{\theta_{ij}}\right)^2 \bigg] \bigg[\prod\limits_{v\in\partial\Delta}
\delta(H_v(G_{ij})) \bigg] \bigg[\prod\limits_i \delta_\star\bigg(\sum\limits_{j> i}\! X_{ij} - \sum\limits_{j<i}
\!\Ad_{G_{ij}}^{-1} X_{ji}\bigg) \bigg]
\nonumber
\\
\hphantom{\tilde{\mathcal{A}}_{\rm PR}^{{\rm lo}}(X_{ij}) \propto}{}
\star \cos\bigg(\frac{i}{\hbar\kappa} \sum\limits_{\substack{i,j
\\
i<j}} |X_{ij}| \theta_{ij}\bigg),
\label{eq:PRtetraregge}
\end{gather}
which is just the formula~\eqref{eq:PRregge} for a~single tetrahedron.
The argument of the cosine in~\eqref{eq:PRtetraregge} is the f\/irst order Regge action for a~tetrahedron.

As the Regge action has previously been recovered in the semi-classical limit of the Ponzano--Regge model in the spin
representation by using coherent states, we wish to link our calculation to the spin basis, which is given by the
${\rm SU}(2)$ Wigner~$D$-matrices $D^{j}_{kl}(g)$.
We f\/ind for the Ponzano--Regge amplitude in the spin basis
\begin{gather}
\tilde{\mathcal{A}}_{\rm PR}(j_{ij};k_{ij},l_{ij}) = \int \bigg[\prod\limits_{i,j} \dd g_{ij}\,
D^{j_{ij}}_{k_{ij}l_{ij}}(g_{ij})\bigg] \mathcal{A}_{\rm PR}(g_{ij})
\nonumber
\\
\hphantom{\tilde{\mathcal{A}}_{\rm PR}(j_{ij};k_{ij},l_{ij}) }{}
= \left\{\begin{matrix} j_{12} & j_{13} & j_{14}
\\
j_{23} & j_{24} & j_{34}
\end{matrix}
\right\} \bigg[\prod\limits_{i<j} \delta^{j_{ij}j_{ji}} \delta_{k_{ij}k_{ji}}\bigg]
 \left(
\begin{matrix} j_{21} & j_{31} & j_{41}
\\
l_{21} & l_{31} & l_{41}
\end{matrix}
\right) \left(
\begin{matrix} j_{21} & j_{32} & j_{42}
\\
-l_{12} & l_{32} & l_{42}
\end{matrix}
\right)
\nonumber
\\
\hphantom{\tilde{\mathcal{A}}_{\rm PR}(j_{ij};k_{ij},l_{ij}) =}{}
\times \left(
\begin{matrix} j_{31} & j_{32} & j_{43}
\\
-l_{13} & -l_{23} & l_{43}
\end{matrix}
\right) \left(
\begin{matrix} j_{41} & j_{42} & j_{43}
\\
-l_{14} & -l_{24} & -l_{34}
\end{matrix}
\right),
\label{eq:PRspin}
\end{gather}
where we introduced the ${\rm SU}(2)$ $6j$-symbol and the $3jm$-symbol, familiarly denoted as
\begin{gather*}
\left\{
\begin{matrix} j_{1} & j_{2} & j_{3}
\\
j_{4} & j_{5} & j_{6}
\end{matrix}
\right\}
\qquad
\text{and}
\qquad
\left(
\begin{matrix} j_1 & j_2 & j_3
\\
m_1 & m_2 & m_3
\end{matrix}
\right),
\end{gather*}
respectively, which are the basic building blocks of Ponzano--Regge model in the spin representation.
Typically, to study the asymptotics of spin foams one considers the formula for the square of the $6j$-symbol expressed
in terms of integrals over ${\rm SU}(2)$ characters $\chi^j:{\rm SU}(2) \rightarrow \mathbb{C}$ as
\begin{gather}
\label{eq:6j}
\left\{
\begin{matrix} j_{12} & j_{13} & j_{14}
\\
j_{23} & j_{24} & j_{34}
\end{matrix}
\right\}^2 = \int \bigg[\prod\limits_{i,j} \dd g_{ij}\bigg] \bigg[\prod\limits_{i} \dd h_{i}\bigg] \bigg[\prod\limits_{i<j}
\chi^{j_{ij}}\big(g_{ij}h_j^{-1}h_{i}g_{ji}^{-1}\big) \bigg],
\end{gather}
where $g_{ij}$ correspond to boundary connection variables, and $h_i$ correspond to parallel transports from the
boundary triangles to the center of the tetrahedron, as before.
This corresponds to the Ponzano--Regge amplitude for a~tetrahedron with f\/ixed quantized edge lengths, since~by
f\/ixing the boundary connection $g_{ij}$ and summing over all $j_{ij}$ (with weights $(2j_{ij}+1)$) we arrive again
at~\eqref{eq:PRtetra}.
More accurately, from~\eqref{eq:PRspin} we f\/ind
\begin{gather*}
\left\{
\begin{matrix} j_{12} & j_{13} & j_{14}
\\
j_{23} & j_{24} & j_{34}
\end{matrix}
\right\}^2 = \sum\limits_{\substack{j_{ji}
\\
i<j}} \sum\limits_{k_{ij},l_{ij}} \bigg[\prod\limits_{i<j}\delta_{l_{ij}l_{ji}}\bigg]
\tilde{\mathcal{A}}_{\rm PR}(j_{ij};k_{ij},l_{ij}).
\end{gather*}
Thus, denoting still by $\tilde{\mathcal{A}}_{\rm PR}(j_{ij};k_{ij},l_{ij})$ the amplitude, where we have set
$j_{ji}=j_{ij}$, $k_{ij}=k_{ji}$ and $l_{ij}=l_{ji}$, we may write
\begin{gather}
\label{eq:6jspin}
\left\{
\begin{matrix} j_{12} & j_{13} & j_{14}
\\
j_{23} & j_{24} & j_{34}
\end{matrix}
\right\}^2 = \sum\limits_{k_{ij},l_{ij}} \tilde{\mathcal{A}}_{\rm PR}(j_{ij};k_{ij},l_{ij}).
\end{gather}

Let us def\/ine functions
\begin{gather*}
D_{\hat{m}\hat{n}}^{j}(g):= \big\langle j,\hat{m}| D^{j}(g) |j,\hat{n}\big\rangle \equiv \big\langle {\tfrac{1}{2}},\hat{m} | D^{\frac{1}{2}}(g) |{\tfrac{1}{2}},\hat{n}\big\rangle^{2j},
\end{gather*}
where $D^j(g)$ are the {\rm SU}(2) Wigner matrices of spin-$j$ representation, and $|j,\hat{m}\rangle$ are Perelomov coherent
states~\cite{Perelomov} on ${\rm SU}(2)$ labelled by a~representation $j \in \{\frac{n}{2}:n\in \mathbb{N}\}$ and a~unit
vector $\hat{m}\in S^2$.
By applying the decomposition of unity in terms of the Perelomov coherent states
\begin{gather*}
(2j+1)\int_{S^2} \frac{\dd \hat{m}}{4\pi} |j,\hat{m}\rangle \langle j,\hat{m}| = \mathbbm{1}_j,
\end{gather*}
it is easy to show that $D_{\hat{m}\hat{n}}^{j}(g)$ constitute an over-complete basis of functions on ${\rm SU}(2)$, as we
have
\begin{gather*}
\delta(g^{-1}g') = \sum\limits_j (2j+1)^3 \int_{S^2} \frac{\dd \hat{m}}{4\pi} \int_{S^2} \frac{\dd \hat{n}}{4\pi}\,
\overline{D_{\hat{m}\hat{n}}^{j}(g)} D_{\hat{m}\hat{n}}^{j}(g').
\end{gather*}
Similarly, for the ${\rm SU}(2)$ character function we may write
\begin{gather}
\label{eq:coherentcharacter}
\chi^j(g) = \tr (D^j(g)) = (2j+1) \int_{S^2} \frac{\dd \hat{m}}{4\pi} D^j_{\hat{m}\hat{m}}(g).
\end{gather}
In the following, without a~serious danger of confusion, we will denote the spin-$\frac{1}{2}$ representation Wigner
matrices and coherent states simply by $D^{\frac{1}{2}}(g) =: g$ and $|\frac{1}{2},\hat{m}\rangle =: |\hat{m}\rangle$,
respectively, so in our notation $D_{\hat{m}\hat{n}}^{j}(g) \equiv \langle \hat{m}|g|\hat{n}\rangle^{2j}$.
Using~\eqref{eq:coherentcharacter}, the $6j$-symbol~\eqref{eq:6j} may be re-expressed as
\begin{gather}
\left\{
\begin{matrix} j_{12} & j_{13} & j_{14}
\\
j_{23} & j_{24} & j_{34}
\end{matrix}
\right\}^2 = \Bigg[\prod\limits_{i,j} (2j_{ij}+1)^2 \int \frac{\dd \hat{m}_{ij}}{4\pi} \frac{\dd \hat{n}_{ij}}{4\pi}
\Bigg] \hat{\mathcal{A}}_{\rm PR}(j_{ij};\hat{m}_{ij},\hat{n}_{ij}).
\label{eq:6jcoh}
\end{gather}
This is simply the equation~\eqref{eq:6jspin} transformed into the coherent state basis.
Here, $\hat{\mathcal{A}}_{\rm PR}(j_{ij};\hat{m}_{ij}$, $\hat{n}_{ij})$ is indeed the Ponzano--Regge amplitude for a~single
tetrahedron with the coherent state labels as boundary metric data, which may be written as
\begin{gather}
\hat{\mathcal{A}}_{\rm PR}(j_{ij};\hat{m}_{ij},\hat{n}_{ij}) = \int \bigg[\prod\limits_{i,j} \dd g_{ij}\,
D^{j_{ij}}_{\hat{m}_{ij}\hat{n}_{ij}}(g_{ij})\bigg] \mathcal{A}_{\rm PR}(g_{ij})
\nonumber
\\
\hphantom{\hat{\mathcal{A}}_{\rm PR}(j_{ij};\hat{m}_{ij},\hat{n}_{ij})}{}
= \left\{
\begin{matrix} j_{12} & j_{13} & j_{14}
\\
j_{23} & j_{24} & j_{34}
\end{matrix}
\right\} \bigg[\prod\limits_{\substack{i,j
\\
i<j}} \delta_{j_{ij}j_{ji}} \delta^{j_{ij}}_{\hat{m}_{ij},-\hat{m}_{ji}} \bigg] \prod\limits_i \left(
\begin{matrix} j_{ij_1} & j_{ij_2} & j_{ij_3}
\\
\hat{n}_{ij_1} & \hat{n}_{ij_2} & \hat{n}_{ij_3}
\end{matrix}
\right),
\label{eq:Ahat1}
\end{gather}
where we denote $\delta^{j_{ij}}_{\hat{m}_{ij},-\hat{m}_{ji}}:= \langle \hat{m}_{ij}|-\hat{m}_{ji}\rangle^{2j_{ij}}$,
and
\begin{gather*}
\left(
\begin{matrix} j_{1} & j_{2} & j_{3}
\\
\hat{n}_{1} & \hat{n}_{2} & \hat{n}_{3}
\end{matrix}
\right):= \sum\limits_{k_i} \left(
\begin{matrix} j_{1} & j_{2} & j_{3}
\\
k_{1} & k_{2} & k_{3}
\end{matrix}
\right) \langle j_1,k_1|j_1,\hat{n}_1\rangle \langle j_2,k_2|j_2,\hat{n}_2\rangle \langle j_3,k_3|j_3,\hat{n}_3\rangle
\end{gather*}
is the Wigner $3jm$-symbol in the coherent state basis.

On the other hand, we may write the same amplitude~\eqref{eq:Ahat1}, transformed from the non-commutative variables, as
\begin{gather}
\hat{\mathcal{A}}_{\rm PR}(j_{ij};\hat{m}_{ij},\hat{n}_{ij}) = \int \bigg[\prod\limits_{\substack{i,j
\\
i<j}} \frac{\dd X_{ij}}{(2\pi\hbar)^3} \widetilde{D^{j_{ij}}_{\hat{m}_{ij}\hat{n}_{ij}}}(X_{ij}) \bigg] \star
\tilde{\mathcal{A}}_{\rm PR}(X_{ij}),
\label{eq:Ahat2}
\end{gather}
where, adopting for convenience the non-commutative structure associated to the symmetric quantization map, we denote~by
\begin{gather}
\widetilde{D^{j}_{\hat{m}\hat{n}}}(X):= \int \frac{\dd g}{\kappa^3} \overline{E(g,X)} D^{j}_{\hat{m}\hat{n}}(g) = \int
\frac{\dd g}{\kappa^3}\exp\left\{\frac{i}{\hbar} \left[-2i\hbar j \ln \langle \hat{m}|g|\hat{n}\rangle
-\zeta_{\tinyS}(g)\cdot X \right] \right\}
\label{eq:Dhat}
\end{gather}
the non-commutative Fourier transform of $D^{j}_{\hat{m}\hat{n}}(g)$.
Now, let us consider the stationary phase approximation of the expression~\eqref{eq:Ahat2}.
The exponential in~\eqref{eq:Dhat} has real and imaginary parts, both of which must be taken into account in the
stationary phase approximation.
For the non-commutative stationary phase equations of this expression in the classical limit $\hbar \rightarrow 0$,
$j\rightarrow\infty$, $\hbar j=\const$, we obtain by a~straightforward calculation $\hat{n} = \Ad_g\hat{m}$ and
$2\hbar j\hat{m} = \frac{1}{\kappa}X$.
Note that, since we had already understood the~$X$ variables as the classical discrete BF variables, this result
conf\/irms that the coherent state variables acquire the correct geometric interpretation in the classical limit.

Then, from~\eqref{eq:PRtetrag} and $2\hbar\kappa j_{ij}\hat{m}_{ij}=X_{ij}$, $\hat{n}_{ij}=\Ad_{g_{ij}}\hat{m}_{ij}$, we
have for the classical limit of~\eqref{eq:Ahat2} the expression
\begin{gather*}
\hat{\mathcal{A}}_{\rm PR}(j_{ij};\hat{m}_{ij},\hat{n}_{ij}) \propto \int \bigg[\prod\limits_{i,j} \dd g_{ij}\bigg]
\bigg[\prod\limits_{v\in\partial\Delta} \delta(H_v(g_{ij})) \bigg] \\
\hphantom{\hat{\mathcal{A}}_{\rm PR}(j_{ij};\hat{m}_{ij},\hat{n}_{ij}) \propto}{}
\times
\bigg[\prod\limits_i \delta\bigg(\sum\limits_{j> i}
j_{ij}\hat{m}_{ij} - \sum\limits_{j<i} \Ad_{g_{ij}^{-1}g_{ji}} j_{ji}\hat{m}_{ji}\bigg) \bigg]
\\
\hphantom{\hat{\mathcal{A}}_{\rm PR}(j_{ij};\hat{m}_{ij},\hat{n}_{ij}) \propto}{}
\times \bigg[\prod\limits_{\substack{(i,j)\in\mathcal{N}
\\
i<j}} \delta(\hat{n}_{ij} - \Ad_{g_{ij}}\hat{m}_{ij}) \bigg] \exp\bigg\{\frac{i}{\hbar} \sum\limits_{\substack{i,j
\\
i<j}} 2\hbar j_{ij}\hat{m}_{ij} \cdot \zeta_{\tinyS}\big(g_{ij}^{-1}g_{ji}\big)\bigg\}\\
\hphantom{\hat{\mathcal{A}}_{\rm PR}(j_{ij};\hat{m}_{ij},\hat{n}_{ij}) \propto}{}
\times (1+\mathcal{O}(\hbar)),
\end{gather*}
where the variation with respect to $X_{ij}$ gives the identif\/ication of the group elements of the two asymptotic
expressions.
Equating the above with the classical limit of the expression~\eqref{eq:Ahat1} and integrating on both sides over all
$\hat{m}_{ij}$, $\hat{n}_{ij}$ as in~\eqref{eq:6jcoh} yields
\begin{gather*}
\left\{
\begin{matrix} j_{12} & j_{13} & j_{14}
\\
j_{23} & j_{24} & j_{34}
\end{matrix}
\right\}^2 \propto \int \bigg[\prod\limits_{\substack{(i,j)\in\mathcal{N}
\\
i<j}} \dd \hat{m}_{ij} \dd G_{ij}\bigg] \exp\bigg\{\frac{i}{\hbar}\sum\limits_{\substack{(i,j)\in\mathcal{N}
\\
i<j}} 2\hbar j_{ij}\hat{m}_{ij}\cdot \zeta_{\tinyS}(G_{ij}) \bigg\} \big(1 + \mathcal{O}(\hbar)\big),
\end{gather*}
where the integral is over sets of unit vectors $\{\hat{m}_{ij}\}$ such that the edge vectors $2\hbar\kappa
j_{ij}\hat{m}_{ij}$ satisfy the closure constraints for each triangle $f_i \in \partial\Delta$ up to parallel transports
given by $G_{ij}$.
As above leading to~\eqref{eq:PRtetraregge} and the f\/irst order Regge action, we may further decompose the integrals
over $G_{ij}$ into integrals over dihedral class angles and unit vectors.
The stationary phase of the exponential of the above amplitude in this limit is again given by the f\/irst order Regge
action
\begin{gather*}
S_{\textrm{\tiny Regge}} = \sum\limits_{\substack{(i,j)\in\mathcal{N}
\\
i<j}} 2\hbar j_{ij}\theta_{ij},
\qquad
\theta_{ij} \in (-\pi,\pi),
\end{gather*}
but now in terms of the coherent state variables.
In essence, the result follows simply because in the classical limit the combination $2\hbar\kappa j_{ij}\hat{m}_{ij}$
of coherent state variables gets identif\/ied with the true phase space variables, the edge vectors $X_{ij}$.
This identif\/ication is due to the asymptotic behavior of the function $\widetilde{D^{j}_{\hat{m}\hat{n}}}(X)$
def\/ined in~\eqref{eq:Dhat}, which mediates the transformation between the coherent state basis and the non-commutative
basis.

\section{Conclusions and comments}
\label{sec:cc}

Let us then summarize our results.
We applied the non-commutative Fourier transform to express the Ponzano--Regge spin foam amplitude as a~f\/irst order
phase space path integral, which took the form of a~discrete BF theory with standard classical action in terms of
non-commutative metric boundary data.
The choice of the quantization map for the geometric observables was seen to be intimately connected to the choice of
a~discretization for the 3d BF theory.
The path integral reformulation then allowed us to study conveniently the classical approximation to the full amplitude.

We discovered that depending on the choice of a~non-commutative structure arising from the deformation quantization
applied to the geometric observables, dif\/ferent limiting behaviors appear for the boundary data in the classical
limit, when we apply the ordinary `commutative' variational calculus to f\/ind the stationary phase solutions.
Furthermore, in this case, the constraints that arise as the classical equations of motion generically do not correspond
to discrete geometries, since the edge vectors in the constraint equations are deformed due to the non-linearity of the
group manifold.
We verif\/ied our observation by considering as explicit examples the non-commutative structures that arise from
symmetric, Duf\/lo and Freidel--Livine--Majid quantization maps.

Accordingly, we were led to consider a~{\it non-commutative} variational method to extract the stationary phase
behavior, which was motivated by the fact that the amplitude for non-commutative metric boundary data acts as the
integral kernel in the propagator with respect to the corresponding $\star$-product, and not the commutative product.
We showed that the non-commutative variations produce the correct geometric constraints for the discrete metric boundary
data in the classical limit.
Thus, we concluded that only by taking into account the deformation of phase space structure in studying the variations,
we f\/ind the undeformed and unambiguous geometric constraints, independent on the choice of the quantization map.

Finally, we considered the asymptotics of the ${\rm SU}(2)$ $6j$-symbol, which is related to the Ponzano--Regge amplitude for
a~tetrahedra with f\/ixed quantized edge lengths.
We found the Regge action, previously recovered in the large spin limit of the $6j$-symbol, in the classical limit.
Our calculations thus not only verify the previous results, but also allows for a~better understanding of them due to
the clear-cut connection to the phase space of classical discretized 3d gravity.
This concrete example also illustrates the use of the non-commutative path integral as a~computational tool.
On the other hand, the full agreement of the results obtained via the non-commutative method with those obtained via
ordinary commutative stationary phase method in the coherent state representation further validates the use of the
non-commutative variations in extracting the asymptotic behavior of the amplitude in the classical limit.

There are several conclusions and further directions of research pointed to by our results.
First and foremost, we have seen that the non-commutative metric representation obtained through the non-commutative
Fourier transform facilitates a~full asymptotic analysis of spin foam models, when proper care is taken in applying
variational methods to the f\/irst order path integral.
In particular, by studying the non-commutative variations one may recover the classical geometric constraints for all
cases of non-commutative structures.
The need for a~non-commutative variational method requires further analysis, and must be taken into account in any
future application of the non-commutative methods to spin foam models.
Our consideration of the $6j$-symbol asymptotics further illustrates the usefulness of the non-commutative methods.

As the non-commutative Fourier transform formalism has recently been extended to all exponential Lie groups~\cite{GOR},
in particular the double-cover ${\rm SL}(2,\mathbb{C})$ of the Lorentz group, we look forward to extending the asymptotic
analysis to the 4d spin foam models in future work, now equipped with the improved understanding of the methods
involved.
In particular, it will be interesting to see how the simplicity constraints turn out to be imposed on the
non-commutative metric variables in the phase space path integral measure for the dif\/ferent spin foam models proposed
in the 4-dimensional case.
The current formulation of 4d models uses the non-commutative Fourier transform for the BF variables (based on ${\rm SO}(4)$,
but extendable to Lorentzian models with ${\rm SL}(2,\mathbb{C})$), on which the simplicity constraints are imposed.
The advantages of the non-commutative formulation are twofold: First, the simplicity constraints may be imposed in
a~very geometrically transparent and natural way via insertions of non-commutative delta functions in the amplitudes, as
one would expect.
Secondly, in such variables, the spin foam amplitudes take again, as in the 3d case, the explicit form of simplicial
gravity path integrals.
As we have shown in this paper, this will greatly facilitate the semi-classical analysis, which would proceed in
entirely the same fashion as the one performed here.
Such an analysis may help to elucidate the dif\/ferences between the geometric properties of the current 4d models, and
even to propose new models with improved geometric behavior in the semi-classical limit.

\appendix

\section{Inf\/initesimal Baker--Campbell--Hausdorf\/f formula}
\label{sec:infBCH}
In order to calculate the deformation matrix~\eqref{eq:DS} for the symmetric quantization map, we need to compute Lie
derivatives $\tilde{\mathcal{L}}_k\zeta_{\tinyS,l}$ of the coordinates $\zeta_{\tinyS} = -\frac{i}{\kappa}\ln(g)$ on
${\rm SU}(2)$.
To do this, we derive the explicit form of the Baker--Campbell--Hausdorf\/f formula $B(k,k')$ in
$e^{ik\cdot\vec{\sigma}} e^{ik'\cdot\vec{\sigma}} \equiv e^{iB(k,k')\cdot\vec{\sigma}}$ for the case, when one of the
arguments is inf\/initesimal.
We may write
\begin{gather*}
\cos|B(k,k')|\mathbbm{1} + i\frac{\sin|B(k,k')|}{|B(k,k')|} B(k,k') \cdot \vec{\sigma}
\nonumber
\\
\qquad{}= \left(\cos|k|\mathbbm{1} + i\frac{\sin|k|}{|k|} k\cdot \vec{\sigma} \right) \left(\cos|k'|\mathbbm{1} +
i\frac{\sin|k'|}{|k'|} k'\cdot \vec{\sigma} \right)
\\
\qquad {}= \left(\cos|k| \cos|k'| - \frac{\sin|k|}{|k|} \frac{\sin|k'|}{|k'|} (k\cdot k') \right) \mathbbm{1}
\\
\qquad\quad{}
+ i \left(\cos|k'| \frac{\sin|k|}{|k|} k + \cos|k| \frac{\sin|k'|}{|k'|} k' - \frac{\sin|k|}{|k|} \frac{\sin|k'|}{|k'|}
(k \wedge k') \right)\cdot \vec{\sigma},
\end{gather*}
where by $\wedge$ we denote the cross-product in $\mathbb{R}^3$, from which one can extract
\begin{gather*}
\cos|B(k,k')| = \cos|k| \cos|k'| - \frac{\sin|k|}{|k|} \frac{\sin|k'|}{|k'|} (k\cdot k'),
\\
\frac{\sin|B(k,k')|}{|B(k,k')|} B(k,k') = \cos|k'| \frac{\sin|k|}{|k|} k + \cos|k| \frac{\sin|k'|}{|k'|} k' -
\frac{\sin|k|}{|k|} \frac{\sin|k'|}{|k'|} (k \wedge k').
\end{gather*}
From these identities it is not too dif\/f\/icult to f\/ind a~closed form for the Baker--Campbell--Hausdorf\/f formula
for ${\rm SU}(2)$ in terms of elementary functions~\cite{GOR}.
However, we will only need to consider the special case $k' = t e_k$, where $t>0$ is an expansion parameter, and $e_k$,
$k=1,2,3$, are orthonormal basis vectors in $\mathbb{R}^3$.
Then we obtain the deformation matrix as $D^{\tinyS}_{kl}(g) = \tilde{\mathcal{L}}_k\zeta_{\tinyS,l}(g) =
\left.
\frac{\dd}{\dd t} B(k(g),te_k)_l\right|_{t=0}$, i.e., it is the~$t$-linear term in $B(k(g),te_k)_l$.
From above we have
\begin{gather*}
\frac{\sin|B(k,te_k)|}{|B(k,te_k)|} B(k,te_k)_l = \frac{\sin|k|}{|k|}k_l + t \left(\cos|k|\delta_{kl} - (k \wedge e_k)_l
\right) + \mathcal{O}\big(t^2\big),
\end{gather*}
from which we may deduce
\begin{gather*}
|B(k,te_k)| = |k| + t\frac{k_k}{|k|} + \mathcal{O}\big(t^2\big),
\\
\frac{1}{\sin|B(k,te_k)|} = \frac{1}{\sin|k|}\left(1 - t \frac{\cos|k|}{\sin|k|}\frac{k_k}{|k|} \right) +
\mathcal{O}\big(t^2\big).
\end{gather*}
Using these formulae, we get
\begin{gather*}
B(k,te_k)_l = k_l + t \frac{|k|}{\sin|k|} \left[\cos|k|\delta_{kl} + \left(\frac{\sin|k|}{|k|} - \cos|k| \right)
\frac{k_k k_l}{|k|^2} - \epsilon_{kl}^{\phantom{kl}m}k_m \right] + \mathcal{O}\big(t^2\big),
\end{gather*}
and so
\begin{gather}
\label{eq:BCHder}
D^{\tinyS}_{kl}(g) = \frac{|k(g)|}{\sin|k(g)|} \!\left[\cos|k(g)|\delta_{kl} + \left(\frac{\sin|k(g)|}{|k(g)|} -
\cos|k(g)| \right)\! \frac{k_k(g) k_l(g)}{|k(g)|^2} - \epsilon_{kl}^{\phantom{kl}m}k_m(g) \right]\!.\!\!\!
\end{gather}

\subsection*{Acknowledgments}

We are grateful for the anonymous referees for their constructive questions and comments, which led to several
improvements to the original manuscript.
We would like to thank A.~Baratin for several useful discussions on the non-commutative Fourier transform and spin foam
models.
We also thank C.~Guedes, F.~Hellmann and W.~Kaminski for several discussions.
This work was supported by the A.~von Humboldt Stiftung, through a~Sofja Kovalevskaja Prize, which is gratefully
acknowledged.
The work of M.~Raasakka was partially supported by Emil Aaltonen Foundation.

\pdfbookmark[1]{References}{ref}
\LastPageEnding


\begin{thebibliography}{99}
\footnotesize \itemsep=0pt

\bibitem{AGN}
Alexandrov S., Geiller M., Noui K., Spin foams and canonical quantization,
  \href{http://dx.doi.org/10.3842/SIGMA.2012.055}{\textit{SIGMA}} \textbf{8} (2012), 055, 79~pages, \href{http://arxiv.org/abs/1112.1961}{arXiv:1112.1961}.

\bibitem{B}
Baez J.C., An introduction to spin foam models of {$BF$} theory and quantum
  gravity, in Geometry and Quantum Physics ({S}chladming, 1999),
  \href{http://dx.doi.org/10.1007/3-540-46552-9_2}{\textit{Lecture Notes in Phys.}}, Vol.~543, Springer, Berlin, 2000, 25--93,
  \href{http://arxiv.org/abs/#2}{gr-qc/9905087}.

\bibitem{Bahr:2009ku}
Bahr B., Dittrich B., ({B}roken) gauge symmetries and constraints in {R}egge
  calculus, \href{http://dx.doi.org/10.1088/0264-9381/26/22/225011}{\textit{Classical Quantum Gravity}} \textbf{26} (2009), 225011,
  34~pages, \href{http://arxiv.org/abs/0905.1670}{arXiv:0905.1670}.

\bibitem{BDOT}
Baratin A., Dittrich B., Oriti D., Tambornino J., Non-commutative f\/lux
  representation for loop quantum gravity, \href{http://dx.doi.org/10.1088/0264-9381/28/17/175011}{\textit{Classical Quantum Gravity}}
  \textbf{28} (2011), 175011, 19~pages, \href{http://arxiv.org/abs/1004.3450}{arXiv:1004.3450}.

\bibitem{BGO}
Baratin A., Girelli F., Oriti D., Dif\/feomorphisms in group f\/ield theories,
  \href{http://dx.doi.org/10.1103/PhysRevD.83.104051}{\textit{Phys. Rev.~D}} \textbf{83} (2011), 104051, 22~pages,
  \href{http://arxiv.org/abs/1101.0590}{arXiv:1101.0590}.

\bibitem{BO}
Baratin A., Oriti D., Group f\/ield theory with noncommutative metric variables,
  \href{http://dx.doi.org/10.1103/PhysRevLett.105.221302}{\textit{Phys. Rev. Lett.}} \textbf{105} (2010), 221302, 4~pages,
  \href{http://arxiv.org/abs/1002.4723}{arXiv:1002.4723}.

\bibitem{BO2}
Baratin A., Oriti D., Quantum simplicial geometry in the group f\/ield theory
  formalism: reconsidering the Barrett--Crane model, \href{http://dx.doi.org/10.1088/1367-2630/13/12/125011}{\textit{New~J. Phys.}}
  \textbf{13} (2011), 125011, 28~pages, \href{http://arxiv.org/abs/1108.1178}{arXiv:1108.1178}.

\bibitem{BO3}
Baratin A., Oriti D., Group f\/ield theory and simplicial gravity path integrals:
  a model for Holst--Pleba\'{n}ski gravity, \href{http://dx.doi.org/10.1103/PhysRevD.85.044003}{\textit{Phys. Rev.~D}} \textbf{85}
  (2012), 044003, 15~pages, \href{http://arxiv.org/abs/1111.5842}{arXiv:1111.5842}.

\bibitem{BarrettCrane}
Barrett J.W., Crane L., Relativistic spin networks and quantum gravity,
  \href{http://dx.doi.org/10.1063/1.532254}{\textit{J.~Math. Phys.}} \textbf{39} (1998), 3296--3302,
  \href{http://arxiv.org/abs/gr-qc/9709028}{gr-qc/9709028},.

\bibitem{BDFHP}
Barrett J.W., Dowdall R.J., Fairbairn W.J., Hellmann F., Pereira R., Lorentzian
  spin foam amplitudes: graphical calculus and asymptotics, \href{http://dx.doi.org/10.1088/0264-9381/27/16/165009}{\textit{Classical
  Quantum Gravity}} \textbf{27} (2010), 165009, 34~pages, \href{http://arxiv.org/abs/0907.2440}{arXiv:0907.2440}.

\bibitem{Barrett08}
Barrett J.W., Naish-Guzman I., The Ponzano--Regge model, \href{http://dx.doi.org/10.1088/0264-9381/26/15/155014}{\textit{Classical
  Quantum Gravity}} \textbf{26} (2011), 155014, 48~pages, \href{http://arxiv.org/abs/0803.3319}{arXiv:0803.3319}.

\bibitem{Boulatov92}
Boulatov D.V., A model of three-dimensional lattice gravity, \href{http://dx.doi.org/10.1142/S0217732392001324}{\textit{Modern
  Phys. Lett.~A}} \textbf{7} (1992), 1629--1646, \href{http://arxiv.org/abs/hep-th/9202074}{hep-th/9202074}.

\bibitem{Caselle89}
Caselle M., D'Adda A., Magnea L., Regge calculus as a local theory of the
  {P}oincar\'e group, \href{http://dx.doi.org/10.1016/0370-2693(89)90441-3}{\textit{Phys. Lett.~B}} \textbf{232} (1989), 457--461.

\bibitem{CD}
Chaichian M., Demichev A., Path integrals in physics. {V}ol.~{I}. Stochastic
  processes and quantum mechanics, \textit{Series in Mathematical and Computational
  Physics}, Institute of Physics Publishing, Bristol, 2001.

\bibitem{CF}
Conrady F., Freidel L., Semiclassical limit of 4-dimensional spin foam models,
  \href{http://dx.doi.org/10.1103/PhysRevD.78.104023}{\textit{Phys. Rev.~D}} \textbf{78} (2008), 104023, 18~pages,
  \href{http://arxiv.org/abs/0809.2280}{arXiv:0809.2280}.

\bibitem{DGO}
Dittrich B., Guedes C., Oriti D., On the space of generalized f\/luxes for loop
  quantum gravity, \href{http://dx.doi.org/10.1088/0264-9381/30/5/055008}{\textit{Classical Quantum Gravity}} \textbf{30} (2013),
  055008, 24~pages, \href{http://arxiv.org/abs/1205.6166}{arXiv:1205.6166}.

\bibitem{DGH}
Dowdall R.J., Gomes H., Hellmann F., Asymptotic analysis of the
  {P}onzano-{R}egge model for handlebodies, \href{http://dx.doi.org/10.1088/1751-8113/43/11/115203}{\textit{J.~Phys.~A: Math. Theor.}}
  \textbf{43} (2010), 115203, 27~pages, \href{http://arxiv.org/abs/0909.2027}{arXiv:0909.2027}.

\bibitem{DGL}
Dupuis M., Girelli F., Livine E., Spinors and {V}oros star-product for group
  f\/ield theory: f\/irst contact, \href{http://dx.doi.org/10.1103/PhysRevD.86.105034}{\textit{Phys. Rev.~D}} \textbf{86} (2012),
  105034, 18~pages, \href{http://arxiv.org/abs/1107.5693}{arXiv:1107.5693}.

\bibitem{EteraMaite}
Dupuis M., Livine E.R., Holomorphic simplicity constraints for 4{D} spinfoam
  models, \href{http://dx.doi.org/10.1088/0264-9381/28/21/215022}{\textit{Classical Quantum Gravity}} \textbf{28} (2011), 215022,
  32~pages, \href{http://arxiv.org/abs/1104.3683}{arXiv:1104.3683}.

\bibitem{Engle07b}
Engle J., Livine E., Pereira R., Rovelli C., L{QG} vertex with f\/inite {I}mmirzi
  parameter, \href{http://dx.doi.org/10.1016/j.nuclphysb.2008.02.018}{\textit{Nuclear Phys.~B}} \textbf{799} (2008), 136--149,
  \href{http://arxiv.org/abs/0711.0146}{arXiv:0711.0146}.

\bibitem{Engle07a}
Engle J., Pereira R., Rovelli C., Loop-quantum-gravity vertex amplitude,
  \href{http://dx.doi.org/10.1103/PhysRevLett.99.161301}{\textit{Phys. Rev. Lett.}} \textbf{99} (2007), 161301, 4~pages,
  \href{http://arxiv.org/abs/0705.2388}{arXiv:0705.2388}.

\bibitem{F}
Freidel L., Group f\/ield theory: an overview, \href{http://dx.doi.org/10.1007/s10773-005-8894-1}{\textit{Internat.~J. Theoret.
  Phys.}} \textbf{44} (2005), 1769--1783, \mbox{\href{http://arxiv.org/abs/hep-th/0505016}{hep-th/0505016}}.

\bibitem{Freidel07}
Freidel L., Krasnov K., A new spin foam model for 4{D} gravity,
  \href{http://dx.doi.org/10.1088/0264-9381/25/12/125018}{\textit{Classical Quantum Gravity}} \textbf{25} (2008), 125018, 36~pages,
  \href{http://arxiv.org/abs/0708.1595}{arXiv:0708.1595}.

\bibitem{FL}
Freidel L., Livine E.R., 3{D} quantum gravity and ef\/fective noncommutative
  quantum f\/ield theory, \href{http://dx.doi.org/10.1103/PhysRevLett.96.221301}{\textit{Phys. Rev. Lett.}} \textbf{96} (2006), 221301,
  4~pages, \href{http://arxiv.org/abs/hep-th/0512113}{hep-th/0512113}.

\bibitem{FM}
Freidel L., Majid S., Noncommutative harmonic analysis, sampling theory and the
  {D}uf\/lo map in {$2+1$} quantum gravity, \href{http://dx.doi.org/10.1088/0264-9381/25/4/045006}{\textit{Classical Quantum Gravity}}
  \textbf{25} (2008), 045006, 37~pages, \href{http://arxiv.org/abs/hep-th/0601004}{hep-th/0601004}.

\bibitem{Goldman1984}
Goldman W.M., The symplectic nature of fundamental groups of surfaces,
  \href{http://dx.doi.org/10.1016/0001-8708(84)90040-9}{\textit{Adv. Math.}} \textbf{54} (1984), 200--225.

\bibitem{GOR}
Guedes C., Oriti D., Raasakka M., Quantization maps, algebra representation,
  and non-commutative {F}ourier transform for {L}ie groups, \href{http://dx.doi.org/10.1063/1.4818638}{\textit{J.~Math.
  Phys.}} \textbf{54} (2013), 083508, 31~pages, \href{http://arxiv.org/abs/1301.7750}{arXiv:1301.7750}.

\bibitem{Han13a}
Han M., On spinfoam models in large spin regime, \href{http://dx.doi.org/10.1088/0264-9381/31/1/015004}{\textit{Classical Quantum
  Gravity}} \textbf{31} (2013), 015004, 21~pages, \href{http://arxiv.org/abs/1304.5627}{arXiv:1304.5627}.

\bibitem{Han13b}
Han M., Semiclassical analysis of spinfoam model with a small
  {B}arbero--{I}mmirzi parameter, \href{http://dx.doi.org/10.1103/PhysRevD.88.044051}{\textit{Phys. Rev.~D}} \textbf{88} (2013),
  044051, 13~pages, \href{http://arxiv.org/abs/1304.5628}{arXiv:1304.5628}.

\bibitem{Han13}
Han M., Krajewski T., Path integral representation of {L}orentzian spinfoam
  model, asymptotics and simplicial geometries, \href{http://dx.doi.org/10.1088/0264-9381/31/1/015009}{\textit{Classical Quantum
  Gravity}} \textbf{31} (2014), 015009, 34~pages, \href{http://arxiv.org/abs/1304.5626}{arXiv:1304.5626}.

\bibitem{HZ1}
Han M., Zhang M., Asymptotics of the spin foam amplitude on simplicial
  manifold: {E}uclidean theory, \href{http://dx.doi.org/10.1088/0264-9381/29/16/165004}{\textit{Classical Quantum Gravity}} \textbf{29}
  (2012), 165004, 40~pages, \href{http://arxiv.org/abs/1109.0500}{arXiv:1109.0500}.

\bibitem{HZ2}
Han M., Zhang M., Asymptotics of spinfoam amplitude on simplicial manifold:
  {L}orentzian theory, \href{http://dx.doi.org/10.1088/0264-9381/30/16/165012}{\textit{Classical Quantum Gravity}} \textbf{30} (2013),
  165012, 57~pages, \href{http://arxiv.org/abs/1109.0499}{arXiv:1109.0499}.

\bibitem{HK}
Hellmann F., Kami\'{n}ski W., Geometric asymptotics for spin foam lattice gauge
  gravity on arbitrary triangulations, \href{http://arxiv.org/abs/1210.5276}{arXiv:1210.5276}.

\bibitem{Hellmann13}
Hellmann F., Kami\'{n}ski W., Holonomy spin foam models: asymptotic geometry of
  the partition function, \href{http://dx.doi.org/10.1007/JHEP10(2013)165}{\textit{J.~High Energy Phys.}} \textbf{2013} (2013),
  no.~10, 165, 63~pages, \href{http://arxiv.org/abs/1307.1679}{arXiv:1307.1679}.

\bibitem{JMN}
Joung E., Mourad J., Noui K., Three dimensional quantum geometry and deformed
  symmetry, \href{http://dx.doi.org/10.1063/1.3131682}{\textit{J.~Math. Phys.}} \textbf{50} (2009), 052503, 29~pages,
  \href{http://arxiv.org/abs/0806.4121}{arXiv:0806.4121}.

\bibitem{KaminskiSteinhaus13}
Kami{\'n}ski W., Steinhaus S., Coherent states, {$6j$} symbols and properties
  of the next to leading order asymptotic expansions, \href{http://dx.doi.org/10.1063/1.4849515}{\textit{J.~Math. Phys.}}
  \textbf{54} (2013), 121703, 58~pages, \href{http://arxiv.org/abs/1307.5432}{arXiv:1307.5432}.

\bibitem{Kawamoto91}
Kawamoto N., Nielsen H.B., Lattice gauge gravity, \href{http://dx.doi.org/10.1103/PhysRevD.43.1150}{\textit{Phys. Rev.~D}}
  \textbf{43} (1991), 1150--1156.

\bibitem{Magliaro11}
Magliaro E., Perini C., Regge gravity from spinfoams, \href{http://dx.doi.org/10.1142/S0218271813500016}{\textit{Internat.~J.
  Modern Phys.~D}} \textbf{22} (2013), 1350001, 21~pages, \href{http://arxiv.org/abs/1105.0216}{arXiv:1105.0216}.

\bibitem{MS}
Majid S., Schroers B.J., {$q$}-deformation and semidualization in 3{D} quantum
  gravity, \href{http://dx.doi.org/10.1088/1751-8113/42/42/425402}{\textit{J.~Phys.~A: Math. Gen.}} \textbf{42} (2009), 425402,
  40~pages, \href{http://arxiv.org/abs/0806.2587}{arXiv:0806.2587}.

\bibitem{Mizoguchi91}
Mizoguchi S., Tada T., Three-dimensional gravity from the {T}uraev--{V}iro
  invariant, \href{http://dx.doi.org/10.1103/PhysRevLett.68.1795}{\textit{Phys. Rev. Lett.}} \textbf{68} (1992), 1795--1798,
  \href{http://arxiv.org/abs/hep-th/9110057}{hep-th/9110057}.

\bibitem{KarimAlex}
Noui K., Perez A., Three-dimensional loop quantum gravity: physical scalar
  product and spin-foam models, \href{http://dx.doi.org/10.1088/0264-9381/22/9/017}{\textit{Classical Quantum Gravity}} \textbf{22}
  (2005), 1739--1761, \href{http://arxiv.org/abs/gr-qc/0402110}{gr-qc/0402110}.

\bibitem{Noui:2011im}
Noui K., Perez A., Pranzetti D., Canonical quantization of non-commutative
  holonomies in {$2+1$} loop quantum gravity, \href{http://dx.doi.org/10.1007/JHEP10(2011)036}{\textit{J.~High Energy Phys.}}
  \textbf{2011} (2011), no.~10, 036, 21~pages, \href{http://arxiv.org/abs/1105.0439}{arXiv:1105.0439}.

\bibitem{Noui:2011aa}
Noui K., Perez A., Pranzetti D., Non-commutative holonomies in $2+1$ LQG and
  Kauf\/fman's brackets, \href{http://dx.doi.org/10.1088/1742-6596/360/1/012040}{\textit{J.~Phys. Conf. Ser.}} \textbf{360} (2012),
  012040, 4~pages, \href{http://arxiv.org/abs/1112.1825}{arXiv:1112.1825}.

\bibitem{O}
Oriti D., The microscopic dynamics of quantum space as a group f\/ield theory, in
  Foundations of Space and Time: Ref\/lections on Quantum Gravity, Editors
  J.~Murugan, A.~Weltman, G.~Ellis, Cambridge University Press, Cambridge,
  2012, 257--320, \href{http://arxiv.org/abs/1110.5606}{arXiv:1110.5606}.

\bibitem{OR}
Oriti D., Raasakka M., Quantum mechanics on {${\rm SO}(3)$} via non-commutative dual
  variables, \href{http://dx.doi.org/10.1103/PhysRevD.84.025003}{\textit{Phys. Rev.~D}} \textbf{84} (2011), 025003, 18~pages,
  \href{http://arxiv.org/abs/1103.2098}{arXiv:1103.2098}.

\bibitem{Perelomov}
Perelomov A., Generalized coherent states and their applications, \href{http://dx.doi.org/10.1007/978-3-642-61629-7}{\textit{Texts and
  Monographs in Physics}}, Springer-Verlag, Berlin, 1986.

\bibitem{P2}
Perez A., The new spin foam models and quantum gravity, \href{http://dx.doi.org/10.4279/PIP.040004}{\textit{Papers Phys.}}
  \textbf{4} (2012), 040004, 37~pages, \href{http://arxiv.org/abs/1205.0911}{arXiv:1205.0911}.

\bibitem{P1}
Perez A., The spin foam approach to quantum gravity, \href{http://dx.doi.org/10.12942/lrr-2013-3}{\textit{Living Rev.
  Relativ.}} \textbf{16} (2013), 3, 128~pages, \href{http://arxiv.org/abs/1205.2019}{arXiv:1205.2019}.

\bibitem{PonzanoRegge68}
Ponzano G., Regge T., Semiclassical limit of Racah coef\/f\/icients, in
  Spectroscopy and Group Theoretical Methods in Physics, Editor F.~Block, North
  Holland, Amsterdam, 1968, 1--58.

\bibitem{Pranzetti:2014}
Pranzetti D., Turaev--Viro amplitudes from $2+1$ loop quantum gravity,
  \href{http://dx.doi.org/10.1103/PhysRevD.89.084058}{\textit{Phys. Rev.~D}} \textbf{89} (2014), 084058, 14~pages,
  \href{http://arxiv.org/abs/1402.2384}{arXiv:1402.2384}.

\bibitem{Regge}
Regge T., Williams R.M., Discrete structures in gravity, \href{http://dx.doi.org/10.1063/1.533333}{\textit{J.~Math.
  Phys.}} \textbf{41} (2000), 3964--3984, \href{http://arxiv.org/abs/gr-qc/0012035}{gr-qc/0012035}.

\bibitem{Reisenberger96}
Reisenberger M.P., Rovelli C., ``{S}um over surfaces'' form of loop quantum
  gravity, \href{http://dx.doi.org/10.1103/PhysRevD.56.3490}{\textit{Phys. Rev.~D}} \textbf{56} (1997), 3490--3508,
  \href{http://arxiv.org/abs/gr-qc/9612035}{gr-qc/9612035}.

\bibitem{ReshetikhinTuraev}
Reshetikhin N., Turaev V.G., Invariants of {$3$}-manifolds via link polynomials
  and quantum groups, \href{http://dx.doi.org/10.1007/BF01239527}{\textit{Invent. Math.}} \textbf{103} (1991), 547--597.

\bibitem{Rovelli:1993kc}
Rovelli C., Basis of the {P}onzano--{R}egge--{T}uraev--{V}iro--{O}oguri
  quantum-gravity model is the loop representation basis, \href{http://dx.doi.org/10.1103/PhysRevD.48.2702}{\textit{Phys. Rev.~D}}
  \textbf{48} (1993), 2702--2707, \href{http://arxiv.org/abs/hep-th/9304164}{hep-th/9304164}.

\bibitem{Rovellibook}
Rovelli C., Quantum gravity, \href{http://dx.doi.org/10.1017/CBO9780511755804}{\textit{Cambridge Monographs on Mathematical Physics}},
  Cambridge University Press, Cambridge, 2004.

\bibitem{Sahlmann}
Sahlmann H., Thiemann T., Chern--{S}imons theory, {S}tokes' theorem, and the
  {D}uf\/lo map, \href{http://dx.doi.org/10.1016/j.geomphys.2011.02.013}{\textit{J.~Geom. Phys.}} \textbf{61} (2011), 1104--1121,
  \href{http://arxiv.org/abs/1101.1690}{arXiv:1101.1690}.

\bibitem{Sahlmann:2011rv}
Sahlmann H., Thiemann T., Chern--Simons expectation values and quantum horizons
  from loop quantum gravity and the Duf\/lo map, \href{http://dx.doi.org/10.1103/PhysRevLett.108.111303}{\textit{Phys. Rev. Lett.}}
  \textbf{108} (2012), 111303, 5~pages, \href{http://arxiv.org/abs/1109.5793}{arXiv:1109.5793}.

\bibitem{S}
Schroers B.J., Combinatorial quantization of {E}uclidean gravity in three
  dimensions, in Quantization of Singular Symplectic Quotients, \textit{Progr.
  Math.}, Vol. 198, Editors N.~Landsman, M.~Pf\/laum, M.~Schlichenmaier,
  Birkh\"auser, Basel, 2001, 307--327, \href{http://arxiv.org/abs/math.QA/0006228}{math.QA/0006228}.

\bibitem{Sengupta2003}
Sengupta A.N., The volume measure for f\/lat connections as limit of the
  {Y}ang--{M}ills measure, \href{http://dx.doi.org/10.1016/S0393-0440(02)00229-2}{\textit{J.~Geom. Phys.}} \textbf{47} (2003),
  398--426.

\bibitem{Thiemannbook}
Thiemann T., Modern canonical quantum general relativity, \href{http://dx.doi.org/10.1017/CBO9780511755682}{\textit{Cambridge Monographs
  on Mathematical Physics}}, Cambridge University Press, Cambridge, 2007.

\bibitem{TuraevViro}
Turaev V.G., Viro O.Y., State sum invariants of {$3$}-manifolds and quantum
  {$6j$}-symbols, \href{http://dx.doi.org/10.1016/0040-9383(92)90015-A}{\textit{Topology}} \textbf{31} (1992), 865--902.

\bibitem{Witten}
Witten E., Quantum f\/ield theory and the {J}ones polynomial, \href{http://dx.doi.org/10.1007/BF01217730}{\textit{Comm. Math.
  Phys.}} \textbf{121} (1989), 351--399.

\bibitem{Witten1991}
Witten E., On quantum gauge theories in two dimensions, \href{http://dx.doi.org/10.1007/BF02100009}{\textit{Comm. Math.
  Phys.}} \textbf{141} (1991), 153--209.

\end{thebibliography}
\end{document}